\documentclass[aps,twocolumn,showpacs]{revtex4-1}

\usepackage{graphicx}  
\usepackage{subfigure}
\usepackage{multirow}
\usepackage{environ}
\usepackage{fancyhdr}
\usepackage{longtable}
\usepackage{parskip}
\usepackage[T1]{fontenc}
\usepackage{dcolumn}   
\usepackage{url,hyperref}
 \usepackage[table,usenames,dvipsnames]{xcolor}
\hypersetup{colorlinks=true, linkcolor=BrickRed,
  urlcolor=blue!50!black, citecolor=blue!50!black}
\usepackage{bm}        
\usepackage{amsfonts}  
\usepackage{amsmath}   
\usepackage{amssymb}   
\usepackage{bbold}
\setcitestyle{super}

\usepackage[capitalize]{cleveref}
\crefrangelabelformat{equation}{(#3#1#4)$-$(#5#2#6)}
\let\ref\cref

\usepackage{xpatch}
\usepackage{appendix}
\crefrangelabelformat{equation}{(#3#1#4)$-$(#5#2#6)}
\xapptocmd\appendices{%
  \crefalias{section}{appendix}%
}{}{\PatchFailed}

\NewEnviron{myequation}{%
    \begin{equation}
    \scalebox{1.2}{$\BODY$}
    \end{equation}
    }

\newcommand{\pwisein}{\left\{ \begin{array}{ll}}
\newcommand{\pwiseout}{\end{array}\right.}

\makeatletter
\@addtoreset{equation}{myequation}
\makeatother

\makeatletter
\@addtoreset{figure}{myfigure}
\makeatother

\begin{document}

\title{Persistence of an active asymmetric rigid Brownian particle in
  two dimensions}

\author{Anirban Ghosh}
\email{anirbansonapur@gmail.com}
\affiliation{Indian Institute of Science Education and Research Mohali,
  Sec. 81, S.A.S. Nagar, Knowledge City, Manauli, Punjab-140306, India.}

\author{Sudipta Mandal}
\affiliation{Indian Institute of Science Education and Research Mohali,
  Sec. 81, S.A.S. Nagar, Knowledge City, Manauli, Punjab-140306, India.}

\author{Dipanjan Chakraborty }
\email{chakraborty@iisermohali.ac.in}
\affiliation{Indian Institute of Science Education and Research Mohali,
  Sec. 81, S.A.S. Nagar, Knowledge City, Manauli, Punjab-140306, India.}

\date{\today}

\begin{abstract}

  We have studied the persistence probability $p(t)$ of an active
  Brownian particle with shape asymmetry in two dimensions. The
  persistence probability is defined as the the probability of a
  stochastic variable that has not changed it's sign in the fixed
  given time interval. We have investigated two cases- diffusion of a
  free active particle and that of harmonically trapped particle. In
  our earlier work, \emph{Ghosh et. al.}, Journal of Chemical
  Physics, \textbf{152},174901, (2020), we had shown that $p(t)$ can
  be used to determine translational and the rotational diffusion
  constant of an asymetric shape particle.  The method has the
  advantage that the measurement of the roational motion of the
  anisotropic particle is not required. In this paper, we extend the
  study to an active an-isotropic particle and show how the
  persistence probability of an anisotropic particle is modified in
  the presence of a propulsion velocity. Further, we validate our
  analytical expression against the measured persistence probability
  from the numerical simulations of single particle Langevin dynamics
  and test whether the method proposed in our earlier work can
  distinguish between an active and a passive anisotropic particle.
  
\end{abstract}

\maketitle 

\section{Introduction}
Persistence plays a very important role in describing a stochastic
processes in nature, specifically the non-stationary dynamics of the
system. The phenomenon of persistence is typically quantified
through the persistence probability. For the last two decades,this
has attracted quite significant attention in the scientific
community. The persistence probability $p(t)$ of a stochastic variable
is the probability that the variable has not changed its sign up to
time $t$. In a wide range of non-equilibrium systems $p(t)$ is found
to decay algebraically with an exponent $\theta$, $p(t)=t^{-\theta}$,
where $\theta$-is a non-trivial exponent. As the temporal correlation
of the non-Markovian stochastic process is highly non-local in
behaviour, the exact calculation of persistence for even simple
non-Markovian stochastic systems is often very difficult and exact
analytical expression for $p(t)$ exists for very few cases. In spite
of this, analytical and or numerical results for the persistence
probability and the exponent $\theta$ exists for systems such as
Brownian motion and diffusion
process\cite{Majumdar:1999tn,majumdar1996,newman1998,sire2000,chakraborty2007,
  chakraborty2008,chakraborty2012c,chakraborty2012d,chakraborty2009,mori2020},
reaction-diffusion systems\cite{derrida1995,cantrell2020},dynamical
systems \cite{menon2003}, phase ordered
kinetics\cite{Majumdar1996c,majumdar1998,Henkel2009}, fluctuating
interfaces\cite{krug1997,kallabis1999,toroczkai1999,majumdar2001a},
critical dynamics\cite{majumdar1996b,chakraborty2007c}, polymer
dynamics\cite{dean2001a,bhattacharya2007}, financial
markets\cite{ren2005,constantin2005} and many more. Even experimental
results exist for persistence exponent in one-dimensional Ising
model\cite{Yurke1997}, diffusion process \cite{wong2001},fluctuating
steps and interfaces\cite{Efraim:2011ks}. For a more comprehensive
review, we invite the readers to the review by Bray \emph{et. al.}
\cite{bray2013b} and Majumdar \cite{Majumdar:1999tn}.

The route to calculating of the persistence probability is through
the non-stationary two-time correlation function. The Lamperti
transformation converts the non-stationary correlator to a stationary
process. For a Gaussian-Markovian stochastic process, the persistence
probability can be directly calculated using Slepian's theorem
\cite{slepian1962}. In contrast, when the process is Non-Markovian,
$p(t)$ is evaluated either using the Independent Interval
Approximation (IIA) when the density of zero crossings stays finite or
a perturbative expansion.

In our earlier work we investigated the effect of shape asymmetry on
the persistence probability of a Brownian particle.\cite{ghosh2020} We
explicitly showed that the measured persistence probability could
estimate the translational and the rotational diffusion constants of
an asymmetric shape particle. The method has the advantage that the
measurement of the rotational motion of the anisotropic particle is
not required. In this paper, we extend the study to an active
anisotropic particle. Our main interest lies in how the persistence
probability of an anisotropic particle is modified in the presence of
a propulsion velocity and whether the method proposed in our earlier
work can distinguish between an active and a passive anisotropic
particle.

This article is organized as follows: In the \cref{sec:II} we have
presented the results for the two-time correlation function for the
position of a free active Brownian particle with shape asymmetry and
along with that the survival probability has been calculated from the
two-time correlation. In the \cref{III}, we have carried out a
perturbative expansion for the position of an anisotropic active
Brownian particle trapped in a harmonic potential. The two-time
correlation has been calculated using the perturbative expansion
method. Finally persistence probability is constructed from the
two-time correlation function.

\section{Active asymmetric particle in two dimensions}
\label{sec:II}

We consider an self propelled asymmetric particle
with velocity $v_0$ in two dimensions with mobilities
$\Gamma_\parallel$ and $\Gamma_\perp$ along the longer and the shorter
axes of the particle, respectively. We have fixed the body frame $x$
and $y$ directions as the long and the short axis, respectively. The
particle has a single rotational mobility $\Gamma_{\theta}$. The
particle is immersed in a bath of temperature $T$ so that the
translational diffusion coefficients along the two directions are
given by $D_{\parallel}=k_BT\Gamma_{\parallel}$ and
$D_{\perp}=k_BT\Gamma_{\perp}$, and the rotational diffusion constant
is $D_{\theta}=k_BT\Gamma_{\theta}$. At a given time $t$ the particle
can be described by the position vector of its center of mass $\vec{r}(t)$
and the angle $\theta(t)$ between the $x$ axis of the lab-frame and
the long axis of the particle. In this frame, the self-propulsion
speed, which is taken along the long axis of the rod, is given by,
$v_0\bold{\hat{n}}(t)$, where
$\hat{n}(t)\equiv(\cos{\theta}(t),\sin{\theta}(t))$ is a unit vector
along the long axis of the particle. In the body frame, the equations
of motion for the center of mass of the particle take the form
\begin{equation}
    \begin{split}
      &\Gamma_1^{-1}\frac{\partial\Tilde{x}}{\partial t}=F_x\cos\theta(t)+F_y\sin{\theta(t)}+\frac{v_0}{\Gamma_1}+\Tilde{\xi_x}(t)\\
      &\Gamma_2^{-1}\frac{\partial\Tilde{y}}{\partial t}=F_y\cos{\theta(t)}-F_x\sin{\theta(t)}+\Tilde{\xi_y}(t)\\
      &\Gamma_3^{-1}\frac{\partial\theta(t)}{\partial
        t}=\tau+\Tilde{\xi_\theta}(t)
    \end{split}
    \label{1}
\end{equation}
\begin{figure}
    \centering
    \includegraphics[width=0.45\textwidth]{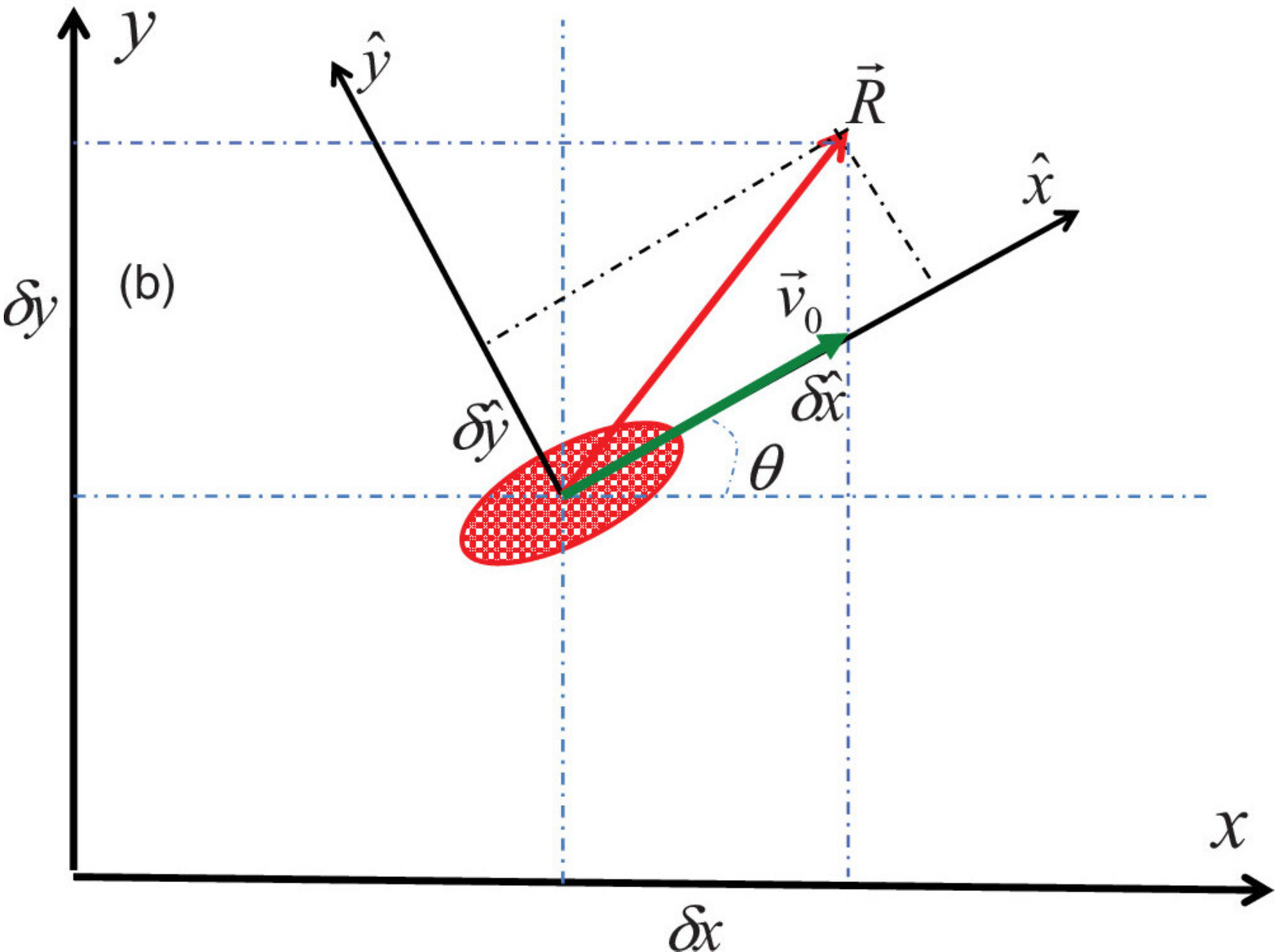}
    \caption{Representation of an ellipsoid in the $x-y$ lab frame and
      the $\hat{x}-\hat{y}$ body frame The angle between two frames is
      $\theta$. The displacement $\Vec{R}$ can be decomposed as
      $(\delta\hat{x},\delta\hat{y})$ or $(\delta x,\delta y)$.}
    \label{fig1}
\end{figure}
Here $F_x$ and $F_y$ are the forces acting on the particle along the
$x$ and $y$ axes(in the lab frame), respectively, and $\tau$ is the
torque acting on the particle. The correlations of the thermal
fluctuations in the body frame are given by

\begin{equation}
    \begin{split}
    &\langle\Tilde{\xi}\rangle=0\\
    &\langle\Tilde{\xi_i}(t)\Tilde{\xi_j}(t^\prime)\rangle=\frac{2k_BT}{\Gamma_i}\delta_{ij}\delta(t-t^\prime)
    \end{split}
    \label{2}
\end{equation}

In the lab frame, the displacements are related to the body frame as

\begin{equation}
    \begin{split}
    &\delta x=\cos{\theta}\delta\Tilde{x}-\sin{\theta}\delta\Tilde{y}\\
    &\delta y=\sin{\theta}\delta\Tilde{x}+\cos{\theta}\delta\Tilde{y}
    \end{split}
    \label{3}
\end{equation}

Using the transformation in \cref{3}, the corresponding equations in
the lab frame is given by

\begin{equation}
    \begin{split}
      &\frac{\partial x}{\partial
        t}=v_0\cos{\theta}(t)+F_x[\Bar{\Gamma}+\frac{\Delta\Gamma}{2}\cos{2\theta}(t)]
      +\frac{\Delta\Gamma}{2} F_y\sin{2\theta}(t)\\
      &+\xi_x(t)\\
      &\frac{\partial y}{\partial
        t}=v_0\sin{\theta}(t)+F_y[\Bar{\Gamma}-\frac{\Delta\Gamma}{2}\cos{2\theta}(t)]
      +\frac{\Delta\Gamma}{2} F_x\sin{2\theta}(t)\\
      &+\xi_y(t)\\
      &\frac{\partial\theta(t)}{\partial t}=\Gamma_3\tau+\xi_\theta(t)
    \end{split}
    \label{4}
\end{equation}

The thermal fluctuations given in \cref{3} are also transformed in the
body frame. The correlations of the thermal fluctuations in the body
frame are given by

\begin{equation}
    \begin{split}
        &\langle\xi_\theta(t)\xi_\theta(t^\prime)\rangle=2D_{\theta}\delta(t-t^\prime)\\
        &\langle\xi_i(t)\xi_j(t^\prime)\rangle_{\theta(t)}^{\xi_1,\xi_2}=2k_BT\Gamma_{ij}\delta(t-t^\prime)
    \end{split}
    \label{6}
\end{equation}

and

\begin{equation}
    \Gamma_{ij}=\Bar{\Gamma}\delta_{ij}+\frac{\Delta\Gamma}{2}\begin{pmatrix}
\cos{2\theta} & \sin{2\theta} \\
\sin{2\theta} & -\cos{2\theta}
\end{pmatrix}
\label{7}
\end{equation}

Here $\Bar{\Gamma}=(\Gamma_\parallel+\Gamma_\perp)/2$ and
$\Delta\Gamma=(\Gamma_\parallel-\Gamma_\perp)$, and mobility tensor
can be written as
$\Gamma_{ij}=\Bar{\Gamma}\delta_{ij}+\frac{\Delta\Gamma}{2}\Delta\mathcal{R}_{ij}[\theta(t)]$,
when the form of $\Delta\Bar{\Bar{\mathcal{R}}}$ is written as

\begin{equation*}
    \Delta\Bar{\Bar{\mathcal{R}}}=\begin{pmatrix}
\cos{2\theta} & \sin{2\theta} \\
\sin{2\theta} & -\cos{2\theta}
\end{pmatrix}
\end{equation*}

\subsection{Mean Square Displacement of the free active particle}
\label{ssec:msd_free}

We first take the case of free active ellipsoidal particle setting the
external potential zero, the equation of motion takes the form

\begin{equation}
    x_i(t)=v_0\int_{0}^{t}\cos{\theta(t^\prime)}dt^\prime+\int_{0}^{t}\xi_i(t^\prime)dt^\prime+x(0)
    \label{8}
\end{equation}

The mean $\langle \Delta x(t) \rangle$, where $\Delta x=x(t)-x(0)$ take
the form  
\begin{equation}
  \label{eq:mean_pos}
  \langle \Delta x(t) \rangle_{\xi,\theta}=v_0 \int_0^t \langle \cos{\theta(t')} \rangle
  \mathrm{d} t'=v_0 \cos \theta_0 \left(\frac{1-e^{-D_\theta t}}{D_\theta}\right).
\end{equation}

The mean square displacement of the particle is calculated from
\cref{8}

\begin{equation}
    \begin{split}
        \langle\Delta
        x_i^2\rangle_{\xi, \theta}&=v_0^2\int_{0}^{t}\langle\cos{\theta}(t^\prime)\cos{\theta}(t^{\prime\prime})\rangle
        dt^\prime dt^{\prime\prime}\\
        &+\int_{0}^{t}\langle\xi_i(t^\prime)\xi_i(t^{\prime\prime})\rangle dt^\prime dt^{\prime\prime}
    \end{split}
    \label{9}
\end{equation}

The explicit evaluation of
the two terms have been shown in \cref{sec:free_particle}. The
final expression for the mean-square displacement, using \cref{A1} and
\cref{A6}, take the form:

\begin{equation}
    \begin{split}
        \langle\Delta
        x^2(t)\rangle_{\xi, \theta}&=2k_BT\Big[\Bar{\Gamma}t+\frac{\Delta\Gamma}{2}\cos{2\theta_0}
        \Big(\frac{1-e^{-4D_\theta t}}{4D_\theta}\Big)\Big]\\
        &+\frac{v_0^2\cos{2\theta_0}}{12D_\theta^2}(3-4e^{-D_\theta t}+e^{-4D_\theta t})\\
        &+\frac{v_0^2}{D_\theta^2}(D_\theta t+e^{-D_\theta t}-1)
    \end{split}
    \label{10}
\end{equation}
and for $y$-direction
\begin{equation}
    \begin{split}
        \langle\Delta
        y^2(t)\rangle_{\xi, \theta}&=2k_BT\Big[\Bar{\Gamma}t-\frac{\Delta\Gamma}{2}\cos{2\theta_0}
        \Big(\frac{1-e^{-4D_\theta t}}{4D_\theta}\Big)\Big]\\
        &-\frac{v_0^2\cos{2\theta_0}}{12D_\theta^2}(3-4e^{-D_\theta t}+e^{-4D_\theta t})\\
        &+\frac{v_0^2}{D_\theta^2}(D_\theta t+e^{-D_\theta t}-1)
    \end{split}
  \end{equation}

In the absence of an active propulsion velocity, the position of the
particle in the lab frame is a non-Gaussian stochastic variable. The
non-Gaussianity parameter is defined as
\begin{equation}
    \phi(t,\theta_0)=\frac{\langle[\Delta x(t)-\langle \Delta x(t)
      \rangle]^4\rangle-3(\langle[\Delta x(t)-\langle \Delta x(t)
      \rangle]^2\rangle)^2}
    {3(\langle[\Delta x(t)-\langle \Delta x(t) \rangle]^2\rangle)^2}
\end{equation}

Defining $\tau_\theta=1/2D_\theta$ and $\tau_n(t)=(1-e^{-nD_\theta
  t})/nD_\theta$, the expressions in \cref{eq:mean_pos} and \cref{10}
take the form 
\begin{equation}
\langle \Delta x(t) \rangle =v_0 \cos \theta_0 \tau_1(t)
\end{equation}

and
\begin{equation}
    \begin{split}
    \langle\Delta x(t)^2\rangle_{\xi, \theta}&=2\bar{D}t+\Delta D\tau_4\cos{2\theta_0}+2\tau_\theta v_0^2\Big((t-\tau_1)\\
    &+\frac{1}{3}(\tau_1-\tau_4)\cos{2\theta_0}\Big)
    \end{split}
\end{equation}
Further, defining $C^{4}_{\theta_0}(t)=\langle[\Delta x(t)-\langle \Delta x(t)
      \rangle_{\xi, \theta}]^4\rangle_{\xi, \theta}-3(\langle[\Delta x(t)-\langle \Delta x(t)
      \rangle_{\xi, \theta}]^2\rangle_{\xi, \theta})^2$, the non-Gaussian parameter is written as, 
\begin{equation}
    \phi(t,\theta_0)=\frac{C^{4}_{\theta_0}(t)}{3(\langle\Delta x(t)^2\rangle)^2}
\end{equation}
\vspace{10mm}
Since we evaluate the persistence probability keeping the initial
angle $\theta_0$ fixed, specifically $\theta_0=0$, we estimate the
non-Gaussian parameter at $\theta_0=0$. Further, we will also consider a
weak asymmetry and weak propulsion velocity so that we evaluate
$\phi(t,\theta_0=0)$ only up to the order of $v_0^2$. The expression
for $C^4_{\theta_0=0}(t)$ takes the form \cite{shee2022}
\begin{widetext}
\begin{equation}
  \label{eq:c4}
  \begin{split}
  C^4_{\theta_0=0}(t)&=\Delta D^2 \Big[\frac{3}{2}t \tau_\theta -3
    \tau^2_4(t)-\frac{1}{2}
    \tau_\theta\tau_{16}(t)-\tau_4(t)\tau_\theta\\
    &+v_0^2\left( 12 \Delta
      D \tau_1(t)\tau_4(t)\Big[1+\frac{16}{36}\tau_\theta\Big]+t\Big[24 \bar{D}
      \tau_1^2(t)+\frac{32}{3}\bar{D}\tau_1(t) \tau_\theta -8 \Delta D
      \tau_4(t)\tau_\theta\Big]-16 \bar{D} \tau_\theta^2 \right)\Big]
  \end{split}
  \end{equation}
\end{widetext}

The expression for $\langle \Delta x^2(t) \rangle_{\xi, \theta}$ up to the order of $v_0^2$ has the
form
\begin{widetext}
\begin{equation}
\begin{split}
  3\langle \Delta x^2(t) \rangle &=12 \bar{D}^2 t^2 +12 \bar{D} \Delta D
  t \tau_4(t) +3 \Delta D \tau_4^2 \\
  &+v_0^2\Big[ \Big(-6 \Delta D \tau_1^2(t)
 \tau_4(t)-8\Delta D \tau_\theta \tau_1(t) \tau_4(t)-4 \Delta D
 \tau_\theta \tau^2_4(t)\Big)\\
 &+ t\Big( -12 \bar{D} \tau^2_1(t) -16
 \bar{D} \tau_1(t) \tau_\theta -8 \bar{D} \tau_4(t) \tau_\theta+12 \Delta
 D   \tau_4(t) \tau_\theta\Big) + 24 \bar{D} \tau_\theta t^2\Big]
\end{split}
\label{eq:xsq_sq}
\end{equation}
\end{widetext}
Clearly from \cref{eq:c4} and \cref{eq:xsq_sq}, the non-Gaussian
parameter depends on the ratio $\Delta D^2/\bar{D}^2$ and
$v^2_0/\bar{D}^2$.In the limit of weak asymmetry and small propulsion
velocity, the non-Gaussian parameter remains small. The time-dependent
$\phi(t,0)$ exhibits a non-monotonic behaviour with a peak at $D_\theta
t \approx 1$.

\subsection{Persistence of the free particle}
We now turn our attention to the persistence probability of a free
asymmetrical active Brownian particle. Setting the external potential
zero, the formal solution to the equation of motion becomes
\begin{equation}
    x_i(t)=x_i(0)+\int_{0}^{t}\xi_i(t^\prime)dt^\prime+v_0\int_{0}^{t}\cos{\theta(t^\prime)}dt^\prime,
    \label{11}
  \end{equation}
  with the initial condition $x_i(0)=0$.
  The calculation of the two-time correlation function
  $\langle x(t_1)x(t_2)\rangle_{\xi_\theta}$ can be achieved by

\begin{equation}
    \begin{split}
      \langle x(t_1)x(t_2)\rangle_{\xi,\theta}&=v_0^2\int
      _{0}^{t_1}dt_1^\prime\int_{0}^{t_2}dt_2^\prime
      \langle\cos{\theta(t_1^\prime)}\cos{\theta(t_2^\prime)}\rangle\\
      &+\int
      _{0}^{t_1}dt_1^\prime\int_{0}^{t_2}dt_2^\prime\langle\xi_1(t_1^\prime)\xi_2(t_2^\prime)\rangle
    \end{split}
    \label{12}
\end{equation}

Considering $t_1>t_2$, the explicit evaluation of the two terms in
\cref{12} has been shown in \cref{ssec:two_time_corr_free}
The final expression is obtained from \cref{A7} and \cref{A11} to give 

\begin{widetext}
\begin{equation}
    \begin{split}
      \langle
      x(t_1)x(t_2)\rangle_{\xi, \theta}&=2k_BT\Bar{\Gamma}t_2\Big[1+\frac{\Delta\Gamma}{2\Bar{\Gamma}}\cos{2\theta_0}
      \Big(\frac{1-e^{-4D_\theta t_2}}{4D_\theta t_2}\Big)\Big]
      +v_0^2\Bigg[\cos{2\theta_0}\Big(\frac{1-e^{-D_\theta t_2}}{6D_\theta^2}+\frac{1-e^{-4D_\theta t_2}}{12D_\theta^2}
      -e^{-D_\theta t_1}\frac{1-e^{-3D_\theta
          t_2}}{6D_\theta^2}\Big)\\
      &-\frac{1-e^{-D_\theta t_2}}{2D_\theta^2}
      +\frac{t_2}{D_\theta} +e^{-D_\theta (t_1-t_2)}\frac{1-e^{-D_\theta t_2}}{D_\theta^2}\Bigg]
    \end{split}
    \label{13}
\end{equation}
\end{widetext}

We now set the initial angle $\theta_0=0$. The diffusion
coefficients $\bar{D}$ and $\Delta D$ are renormalized by the active
velocity. Furthermore, we note that the last term in \cref{13}
contains a stationary component which survives in the long time limit
of $t_1$ and $t_2$ large but $(t_1-t_2)$ finite. This, of course, makes
the coversion of this non-stationary correlator to a stationary one
slightly problematic. In order to transform the non-stationary
correlation into a stationary correlator, we make the approximation
$t_1>>t_2$ so that both the terms
$2 v_0^2 \tau_\theta \tau_3(t) e^{-D_\theta t_1}$ and the last term
$v_0^2e^{-D_\theta t_1} (1-e^{D_\theta t_2})/D_\theta^2$ in \cref{13}
can be dropped.
\begin{equation}
    \begin{split}
      \langle
      x(t_1)x(t_2)\rangle_{\xi, \theta}&=\Big(2k_BT\Bar{\Gamma}+v_0^2/D_\theta\Big)
      t_2\\
      &+ \Big(\Delta D +\frac{v_0^2}{3D_\theta}\Big)\Big(\frac{1-e^{-4D_\theta t_2}}{4D_\theta }\Big)
      -v_0^2\Big(\frac{1-e^{-D_\theta t_2}}{3D_\theta^2}\Bigg)
    \end{split}
    \label{13a}
\end{equation}

Dropping the second term is strictly valid only when $t_1>>
t_2$. Nevertheless, even with this approximation, we want to figure
out how well the analytical expression for $p(t)$ compares with the
numerical results. We use the Lamperti transformation and define
$\Tilde{X}(t)=x(t)/\sqrt{\langle x^2(t)\rangle_{\xi_\theta}}$. The
two-time correlation function of the rescaled variable
$\langle\Tilde{X}(t_1)\Tilde{X}(t_2)\rangle_{\xi_\theta}$ becomes
\begin{widetext}
  \begin{equation}
    \begin{split}
      \langle\Bar{X}(t_1)\Bar{X}(t_2)\rangle=(t_2/t_1)^{1/2}
           \Bigg[2 D_{\rm eff}+\Delta
              D_{\rm eff} \Big(\frac{1-e^{-4D_\theta t_2}}{4D_\theta
                t_2}\Big)
              &-\frac{v_0^2}{6D_\theta}\Big(\frac{1-e^{-D_\theta t_2}}{D_\theta
        t_2}\Big)\Bigg]^{1/2} \\
      &\Bigg[2 D_{\rm eff}+\Delta
              D_{\rm eff}
      \Big(\frac{1-e^{-4D_\theta t_1}}{4D_\theta
        t_1}\Big)-\frac{v_0^2}{6D_\theta}\Big(\frac{1-e^{-D_\theta t_1}}{D_\theta
        t_1}\Big)\Bigg]^{-1/2}
      \label{15}
      \end{split}
  \end{equation}
\end{widetext}
where the effective diffusivity is given by $D_{\rm
  eff}=\bar{D}+v_0^2/2D_\theta$ and $\Delta D_{\rm eff}=\Delta D+v_0^2/3D_\theta$.
We now define the transformation in time as
\begin{widetext}
\begin{equation}
e^T=\sqrt{2D_{\rm eff}t\Bigg[1+\frac{\Delta D_{\rm eff}}{2D_{eff}}
      \Big(\frac{1-e^{-4D_\theta t}}{4D_\theta
        t}\Big)-\frac{v_0^2}{6D_\theta
        D_{\rm eff}}\Big(\frac{1-e^{-D_\theta t}}{D_\theta
        t}\Big)\Bigg]}
        \label{15a}
      \end{equation}
 \end{widetext}

 Using this transformation in time the two-time correlation function
 $\langle \bar{X}(T_1) \bar{X}(T_2)\rangle$ from \cref{15} takes the
 simple form of
 $\langle\Bar{X}(t_1)\Bar{X}(t_2)\rangle=e^{-(T_1-T_2)/2}$. Since the
 stationary correlation function now decays exponentially for all
 times, following Slepian\cite{slepian1962}, the asymptotic form of
 the persistence probability is found as
\begin{equation}
    P(T)=e^{-\lambda T}
    \label{18a}
\end{equation}
    
Transforming back to real-time $t$, we get the persistence
probability for the free particle as
\begin{equation}
    \begin{split}
        p(t,\theta_0=0)
        =\frac{1}{\sqrt{2D_{\rm eff}t}}\Bigg[1&+\frac{\Delta D_{\rm
              eff}}{2D_{\rm eff}}\left(\frac{1-e^{-4D_\theta t}}{4D_\theta
            t}\right)\\
          &-\frac{v_0^2}{6D_\theta D_{\rm eff}}\left(\frac{1-e^{-D_\theta t}}{D_\theta t}\right)\Bigg]^{-1/2}
    \end{split}
    \label{19A}
\end{equation}
Rearranging the above expression, we get
\begin{equation}
    \begin{split}
        t^{1/2}p(t,\theta_0=0)=\frac{1}{\sqrt{2D_{\rm
              eff}}}\Bigg[1&+(\frac{\Delta D_{\rm eff}}{2D_{\rm eff}})
            \left(\frac{1-e^{-4D_\theta t}}{4D_\theta t}\right)\\
            &-\frac{v_0^2}{6D_\theta D_{\rm eff}}\left(\frac{1-e^{-D_\theta t}}{D_\theta t}\right)\Bigg]^{-1/2}
    \end{split}
    \label{19b}
\end{equation}
In the absence of the propulsion velocity $v_0=0$, we recover the
persistence probability of free anisotropic particle.\cite{ghosh2020}

\begin{figure}
    \centering
    \includegraphics[width=0.45\textwidth]{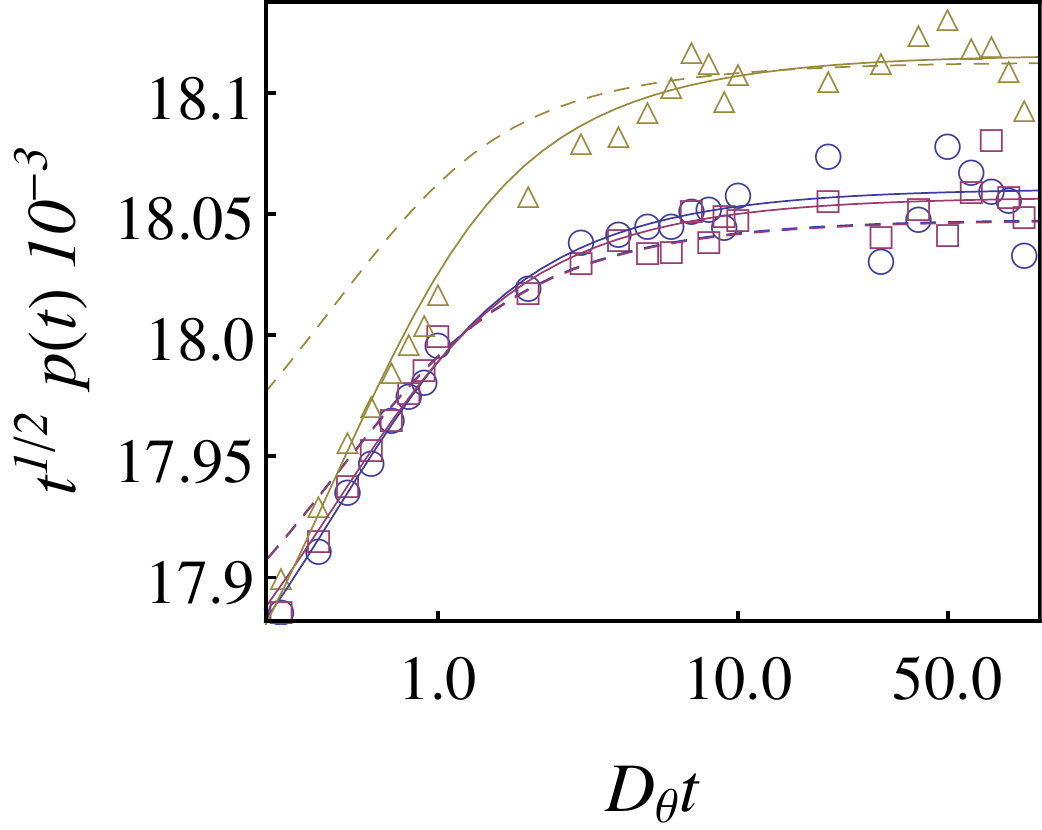}
    \caption{Plot of $t^{1/2}p(t)$ for different choices of propulsion
      velocity $v_0$ of the anisotropic particle: $v_0=0$(open
      circles);$v_0=0.01$(open square) and $v_0=0.1$ (open
      triangles). The translational diffusivities were fixed at
      $D_\parallel=1$, $D_\perp=0.95$. The rotational diffusivity and
      the initial angle $\theta_0$ were fixed at $D_\theta=1$ and
      $\theta_0=0$, respectively. The dashed lines are the plot of
      \cref{19b} whereas the solid lines are fit to the data using
      \cref{19b} with $D_{eff}$ and $\Delta D$ as fit parameters.}
    \label{fig2}
  \end{figure}

\begin{figure}
    \centering
    \includegraphics[width=0.45\textwidth]{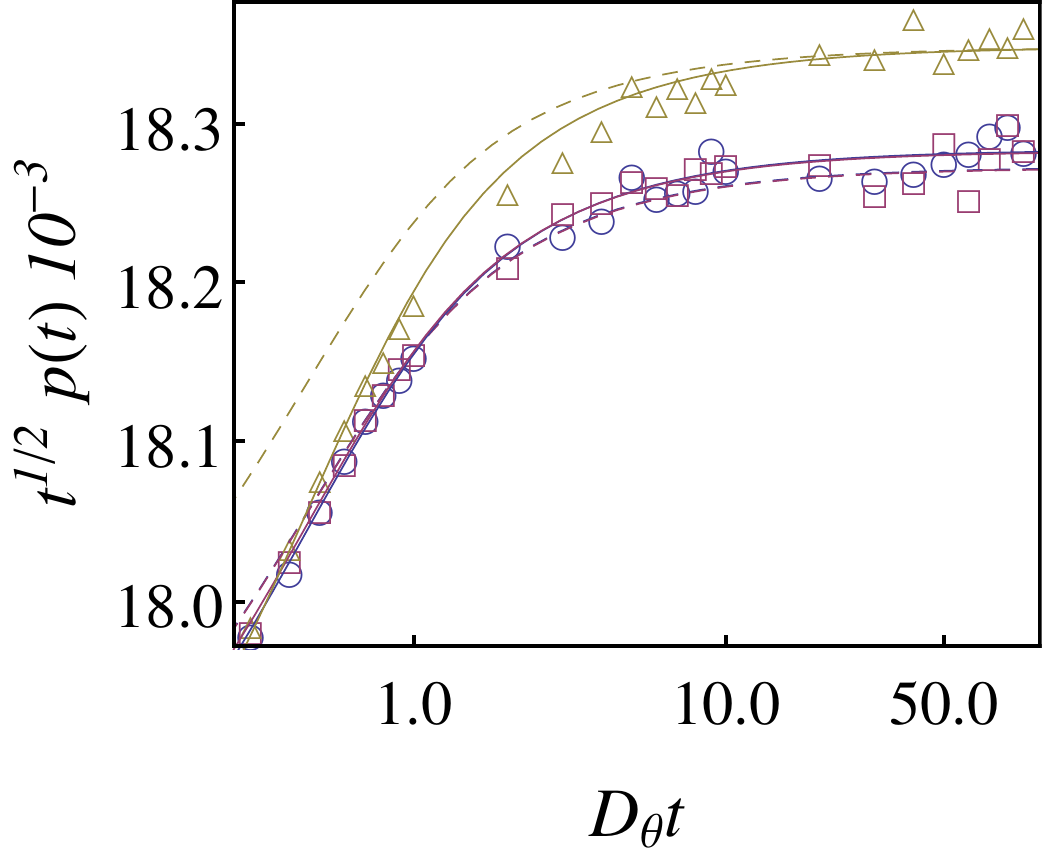}
    \caption{Plot of $t^{1/2}p(t)$ for different choices of propulsion
      velocity $v_0$ of the anisotropic particle: $v_0=0$(open
      circles);$v_0=0.01$(open square) and $v_0=0.1$ (open
      triangles). The translational diffusivities were fixed at
      $D_\parallel=1$, $D_\perp=0.90$. The rotational diffusivity and
      the initial angle were fixed at $D_\theta=1$ and $\theta_0=0$,
      respectively. The dashed lines are the plot of \cref{19b}
      whereas the solid lines are fit to the data using \cref{19b}
      with $D_{eff}$ and $\Delta D$ as fit parameters.}
    \label{fig3}
  \end{figure}

In order to validate the expression for the persistence probability we
performed numerical simulations of \cref{4}. The initial condition was
chosen from a Gaussian distribution with a very small width, so 
the sign of $\vec{r}(0)$ is clearly defined. The trajectories were
evolved in time with an integration time-step of $\delta t=0.001$. At
every instant, the survival of the particle trajectory was checked by
looking at the sign of $\vec{r}(t)$. Fraction of trajectories for
which the position did not change its sign up to time $t$ gave the
survival probability $p(t)$. A total of $10^9$ trajectories were used
in estimating the survival probability. A comparison of the measured
$p(t)$ with that of the predictions of \cref{19b} is shown \cref{fig2}
and \cref{fig3}. Both the figures compare the persistence probability
for weakly asymmetric particles. We observe that while the asymmetry of
the particle is picked up as expected from our earlier work
\cite{ghosh2020}, for small propulsion velocity $t^{1/2}p(t)$ is unable
to pick up the activity of the paricle. In the case when the activity
of the particle is comparatively large, the $t^{1/2}p(t)$ indeed picks
up the activity of the particle. When compared with the analytical
expression of \cref{19b}, for the small propulsion velocity, the
expression compares quite well with the simulation resutls with 
the overall constant as the only fit parameter (the dotted lines in the
figures). When the data is fitted to \cref{19b} with the overall
constant fixed and $\bar{D}$ and $\Delta D$ as fit parameters, it
yeilds the correct values of $\bar{D}$ and $\Delta D$.  However, for
comparatively larger values of $v_0$, when the expression is plotted
against the numerical data with the overall constant as the only fit
parameter, the data matches only asymptotically with the analytical
expression. On the other hand, when the data is fitted with \cref{19b}
with $\bar{D}$ and $\Delta D$ as fit parameters, the fit yeilds a
slightly lower value of $\bar{D}$ and a slightly higher value of
$\Delta D$. For example, in the case of $\bar{D}=0.975$ and
$\Delta D=0.05$ (\cref{fig2} open triangles), we obtain from the fit a
value of $\bar{D}\approx 0.96$ and $\Delta D \approx 0.08$. In the
case of $\bar{D}=0.95$ and $\Delta D=0.1$ (\cref{fig2} open
triangles), we obtain from the fit a value of $\bar{D}\approx 0.93$
and $\Delta D \approx 0.14$.
  
  \section{HARMONICALLY TRAPPED ASYMMETRIC PARTICLE}
  \label{III}

  We now consider the case when such an active anisotropic particle is
  trapped in a harmonic trap. This situation typically occurs in the
  trapping and tracking of colloids in experiments. The harmonic
  potential is taken as isotropic potential with no preferred
  directional alignment.The potential is taken as
  $U(x,y)=\kappa(x^2+y^2)/2$. Using \cref{4} the corresponding
  Langevin equation becomes

\begin{equation}
    \begin{split}
        &\frac{\partial x}{\partial t}=-\kappa
        x(\Bar{\Gamma}+\frac{\Delta\Gamma}{2}\cos{2\theta(t)})
        -\kappa y\frac{\Delta\Gamma}{2}\sin{2\theta(t)}+v_0\cos{\theta(t)}\\
        &+\xi_1(t)\\
        &\frac{\partial y}{\partial t}=-\kappa
        x\frac{\Delta\Gamma}{2}\sin{2\theta(t)}-\kappa y(\Bar{\Gamma}
        -\frac{\Delta\Gamma}{2}\cos{2\theta(t)})+v_0\sin{\theta(t)}\\
        &+\xi_2(t)\\
        &\frac{\partial\theta}{\partial t}=\Gamma_3\tau+\xi_3(t)
    \end{split}
    \label{20}
\end{equation}
The correlations of the thermal fluctuations follow \cref{6}. 

\subsection{Perturbative expansion}
\label{ssec:pertubative_expansion}
Let us define the vector space $\bold{R}\equiv(x,y)^T$, and the
equation takes the general form as
\begin{equation}
  \Dot{\bold{R}}=-\kappa[\Bar{\Gamma}\mathbf{1}+\frac{\Delta\Gamma}{2}\Bar{\Bar{\mathcal{R}}}(t)]\bold{R}(t)
  +v_0\bold{\hat{n}}+\xi(t)
    \label{21}
\end{equation}

In order to solve this equation we use the perturbative expansion

\begin{equation}
  \begin{split}
    \bold{R}(t)=\bold{R}_0(t)-\left(\frac{\kappa\Delta\Gamma}{2}\right)\bold{R}_1(t)
    +\left(\frac{\kappa\Delta\Gamma}{2}\right)^2&\bold{R}_2(t)\\
    &+\mathcal{O}\left(\frac{\kappa\Delta\Gamma}{2}\right)^3
        \end{split}
    \label{22}
\end{equation}

Substituting \cref{22} in \cref{21} and equalizing both
sides we get the equations for $\bold{R}_0(t)$ and $\bold{R}_1(t)$ as

\begin{equation}
    \begin{split}
        &\Dot{\bold{R}}_0(t)=-\kappa\Bar{\Gamma}\bold{R}_0(t)+v_0\bold{\hat{n}}(t)+\xi(t)\\
        &\Dot{\bold{R}}_1(t)=-\kappa\Bar{\Gamma}\bold{R}_1(t)+\Bar{\Bar{\mathcal{R}}}(t)\bold{R}_0(t)\\
        &\Dot{\bold{R}}_2(t)=-\kappa\Bar{\Gamma}\bold{R}_2(t)+\Bar{\Bar{\mathcal{R}}}(t)\bold{R}_1(t)
    \end{split}
    \label{23}
\end{equation}

The formal solutions for \cref{23} with the initial condition $\bold{R}(0)=0$, becomes

\begin{equation}
    \begin{split}
      &\bold{R}_0(t)=\int_{0}^{t}dt^\prime
      e^{-\kappa\Bar{\Gamma}(t-t^\prime)}
      \big[\xi(t^\prime)+v_0\bold{\hat{n}}(t^\prime)\big]\\
      &\bold{R}_1(t)=\int_{0}^{t}dt^\prime
      e^{-\kappa\Bar{\Gamma}(t-t^\prime)}
      \Bar{\Bar{\mathcal{R}}}(t^\prime)\bold{R}_0(t^\prime)\\
      &\bold{R}_2(t)=\int_{0}^{t}dt^\prime
      e^{-\kappa\Bar{\Gamma}(t-t^\prime)}\Bar{\Bar{\mathcal{R}}}(t^\prime)\bold{R}_1(t^\prime)
    \end{split}
    \label{24}
\end{equation}

The explicit form of the correlation matrix $\langle R_i(t)R_j(t) \rangle_{\xi, \theta}$ in the equal time, is given by
\begin{equation}
    \begin{split}
    &\langle R_i(t)R_j(t)\rangle_{\xi,\theta}=\langle
    R_{0,i}(t)R_{o,j}(t)\rangle_{\xi,\theta}\\
    &-\left(\kappa\Delta\Gamma\right)\langle R_{0,i}(t)R_{1,j}\rangle_{\xi,\theta}\\
    &+\left(\frac{\kappa\Delta\Gamma}{2}\right)^2\Big[\langle
    R_{1,i}(t)R_{1,j}(t)
    \rangle_{\xi,\theta}+2\langle
    R_{0,i}(t)R_{2,j}(t)\rangle_{\xi,\theta}\Big]\\
    &+\mathcal{O}\left(\frac{\kappa\Delta\Gamma}{2}\right)^3
    \end{split}
    \label{25}
  \end{equation}
Here we have considered the fact that $\langle R_{0,i}R_{1,j}\rangle=\langle R_{0,j}R_{1,i}\rangle$. 
We now start to calculate the different terms of the correlation
matrix. The correlation matrix for $\bold{R}_0(t)$ averaged over the
translational and the rotational noise is given as
\begin{widetext}
  \begin{equation}
      \langle
      R_{0,i}(t)R_{0,j}(t)\rangle_{\xi, \theta}=\int_{0}^{t}dt^\prime\int_{0}^{t}dt^{\prime\prime}e^{-\kappa\Bar{\Gamma}(t-t^\prime)}
      e^{-\kappa\Bar{\Gamma}(t-t^{\prime\prime})}\langle\xi(t^\prime)\xi(t^{\prime\prime})\rangle
      +\int_{0}^{t}dt^\prime\int_{0}^{t}dt^{\prime\prime}e^{-\kappa\Bar{\Gamma}(t-t^\prime)}e^{-\kappa\Bar{\Gamma}(t-t^{\prime\prime})}
      v_0^2\langle n_i(t^\prime)n_j(t^{\prime\prime})\rangle
    \label{26}
\end{equation}
\end{widetext}
The calculations of the time integrals in \cref{26} have been
explicitly shown in \cref{R0R0} (from \cref{A12} to
\cref{A17}) and the final result for $\langle x_0^2(t) \rangle$ is
found as

\begin{equation}
    \begin{split}
        &\langle
        x_0^2(t)\rangle_{\xi, \theta}=\frac{k_BT}{\kappa}(1-e^{-2\kappa\Bar{\Gamma}t})+\Delta
          D\cos{2\theta_0}
        \Big(\frac{e^{-4D_\theta t}-e^{-2\kappa\Bar{\Gamma}t}}{2\kappa\Bar{\Gamma}-4D_\theta}\Big)\\
        &+\frac{v_0^2\cos{2\theta_0}}{2}\Big[\frac{2D_\theta(e^{-4D_\theta
            t}-e^{-2\kappa\Bar{\Gamma}t})}
        {(2\kappa\Bar{\Gamma}-4D_\theta)(\kappa\Bar{\Gamma}-3D_\theta)(\kappa\Bar{\Gamma}-D_\theta)}\\
        &+\frac{e^{-4D_\theta
            t}-2e^{-(\kappa\Bar{\Gamma}+D_\theta)t}+e^{-2\kappa\Bar{\Gamma}t}}{(\kappa\Bar{\Gamma}-D_\theta)
          (\kappa\Bar{\Gamma}-3D_\theta)}\Big]\\
        &+\frac{v_0^2}{2}\Big[\frac{1-2e^{-(\kappa\Bar{\Gamma}+D_\theta)t}+e^{-2\kappa\Bar{\Gamma}t}}{(\kappa\Bar{\Gamma}-D_\theta)
          (\kappa\Bar{\Gamma}+D_\theta)}
        -\frac{D_\theta(1-e^{-2\kappa\Bar{\Gamma}t})}{\kappa\Bar{\Gamma}(\kappa\Bar{\Gamma}-D_\theta)(\kappa\Bar{\Gamma}+D_\theta)}\Big]
    \end{split}
    \label{27}
\end{equation}

In the limit of $\kappa\rightarrow 0$, \cref{27} reproduces \cref{10},
the correct result for the free diffusion of active anisotropic particles.

We now proceed to calculate the correction term to this correlation
and restrict ourselves only to the first order correction.  To this
end, we first restructure the solution of $\bold{R}_1(t)$ as

\begin{equation}
    R_{1,i}(t)=\int_{0}^{t}dt^\prime e^{-\kappa\Bar{\Gamma}(t-t^\prime)}\sum_j\mathcal{R}_{ij}(t^\prime)R_{0,j}(t^\prime)
    \label{28}
\end{equation}

The correlation function $\langle
R_{0,i}(t)R_{1,j}(t)\rangle_{\xi,\theta}$ is then given by
\begin{align}
  \nonumber
      &\langle R_{0,i}(t_1)R_{1,j}(t_2)\rangle_{\xi, \theta}=
        \begin{aligned}[t]
      \nonumber
      \Bigg\langle
        R_{0,i}(t_1)\int_{0}^{t_2}dt_2^\prime
        &e^{-\kappa\Bar{\Gamma}(t_2-t_2^\prime)}\\
        &\sum_k\mathcal{R}_{jk}(t_2^\prime)R_{0,k}(t_2^\prime)\Bigg\rangle\\
      \end{aligned}
  \nonumber\\
  \nonumber
        &=\Bigg\langle\int_{0}^{t_2}dt_2^\prime
        e^{-\kappa\Bar{\Gamma}(t_2-t_2^\prime)}\sum_k\mathcal{R}_{jk}(t_2^\prime)R_{0,i}(t_1)
        R_{0,k}(t_2^\prime)\Bigg\rangle\\
    \label{28a}
    \end{align}
    We now substitute the formal solution for $R_{0,i}(t)$ as given in
    \cref{24} and solve the integrals over time. The detailed
    calculation of \cref{28a} for the specific term
    $\langle x_0()x_1(t)\rangle$ is given in \cref{subsec:R_0R_1}.
    In evaluating the integrals, we have used the identity 
\begin{equation}
    \langle
    e^{im\Delta\theta(t^\prime)-im\Delta\theta(t^{\prime\prime})}\rangle=e^{-D_\theta\left[m^2t^\prime+n^2t^{\prime\prime}-2mn
      \min(t^\prime, t^{\prime\prime})\right]}
    \label{29}
\end{equation}
Accordingly, the averages of the trigonometric functions over
the rotational noise, which are used in the calculations, take the form
\begin{equation}
    \begin{split}
        &\langle\cos{2[\theta(t^\prime)-\theta(t^{\prime\prime})]}\rangle_{\theta}=e^{-4D_\theta[t^\prime+t^{\prime\prime}-2\min(t^\prime,t^{\prime\prime})]}\\
        &\langle\cos{2[\theta(t^\prime)+\theta(t^{\prime\prime})]}\rangle_{\theta}=\cos{4\theta_0}e^{-4D_\theta[t^\prime+t^{\prime\prime}
          +2\min(t^\prime,t^{\prime\prime})]}\\
        &\langle\sin{2[\theta(t^\prime)+\theta(t^{\prime\prime})]}\rangle_{\theta}=\sin{4\theta_0}
        e^{-4D_\theta[t^\prime+t^{\prime\prime}+2\min(t^\prime,t^{\prime\prime})]}\\
        &\langle\sin{2[\theta(t^\prime)-\theta(t^{\prime\prime})]}\rangle_{\theta}=0
    \end{split}
\end{equation}

The final expression for $\langle x_0(t)x_1(t) \rangle$ is given by
\begin{widetext}
\begin{equation}
\begin{split}
\langle
x_0(t)x_1(t)\rangle_{\xi, \theta}=\Big(\frac{k_BT}{\kappa}\Big)\cos{2\theta_0}
\bigg(\frac{e^{-4D_\theta
    t}-e^{-2\kappa\overline{\Gamma}t}}{2\kappa\overline{\Gamma}-4D_\theta}
&-\frac{e^{-2\kappa\overline{\Gamma}t}-
  e^{-2(2D_\theta+\kappa\overline{\Gamma})t}}{4D_\theta}\bigg)
+\Big(\frac{k_BT}{\kappa}\Big)\Big(\frac{\Delta\Gamma}{2\overline{\Gamma}}\Big)
\bigg[\frac{1-e^{-2\kappa\overline{\Gamma}t}}{2\kappa\overline{\Gamma}+4D_\theta}\\
&-\frac{2\kappa\overline{\Gamma}}{4D_\theta}\frac{e^{-2\kappa\overline{\Gamma}t}-e^{-(2\kappa\overline{\Gamma}+4D_\theta)t}}
{\kappa\overline{\Gamma}+4D_\theta}\bigg]
-\frac{3v_0^2D_\theta
  e^{-\kappa\bar{\Gamma}t}\sinh{\kappa\bar{\Gamma}t}}
{2\kappa\bar{\Gamma}(\kappa\bar{\Gamma}+D_\theta)(\kappa\bar{\Gamma}-D_\theta)(\kappa\bar{\Gamma}+2D_\theta)}
\end{split}
\label{28b}
\end{equation}
\end{widetext}

Following \cref{25}, the mean-square displacement
$\langle x^2(t) \rangle_{\xi, \theta}$ up to the first order correction is given by
$\langle x^2(t) \rangle_{\xi, \theta}=\langle x_0^2(t) \rangle_{\xi,\theta} -(\kappa
\Delta \Gamma) \langle x_0(t)x_1(t) \rangle_{\xi,\theta}$. From
\cref{27,28b} it is clear that the second term in \cref{27} cancels
with the first term in \cref{28b}. Further, since we are interested in
the expression for the mean square displacement up to the first order,
the expression for $\langle x^2(t) \rangle$ becomes
\begin{widetext}
  \begin{equation}
    \begin{split}
\langle x^2(t)\rangle_{\xi, \theta}&=\left(\frac{k_BT}{\kappa}\right)\Bigg[(1-e^{-2\kappa\Bar{\Gamma}t})+\kappa
\Delta \Gamma \cos{2\theta_0} \bigg(\frac{e^{-2\kappa\overline{\Gamma}t}-
  e^{-(4D_\theta+2\kappa\overline{\Gamma})t}}{4D_\theta}\bigg)\\
       & +\frac{v_0^2\cos{2\theta_0}}{2}\Big[\frac{2D_\theta(e^{-4D_\theta
            t}-e^{-2\kappa\Bar{\Gamma}t})}
    {(2\kappa\Bar{\Gamma}-4D_\theta)(\kappa\Bar{\Gamma}-3D_\theta)(\kappa\Bar{\Gamma}-D_\theta)}
    +\frac{e^{-4D_\theta
            t}-2e^{-(\kappa\Bar{\Gamma}+D_\theta)t}+e^{-2\kappa\Bar{\Gamma}t}}{(\kappa\Bar{\Gamma}-D_\theta)
          (\kappa\Bar{\Gamma}-3D_\theta)}\Big]\\
        &+\frac{v_0^2}{2}\Big[\frac{1-2e^{-(\kappa\Bar{\Gamma}+D_\theta)t}+e^{-2\kappa\Bar{\Gamma}t}}{(\kappa\Bar{\Gamma}-D_\theta)
          (\kappa\Bar{\Gamma}+D_\theta)}
        -\frac{D_\theta(1-e^{-2\kappa\Bar{\Gamma}t})}{\kappa\Bar{\Gamma}(\kappa\Bar{\Gamma}-D_\theta)(\kappa\Bar{\Gamma}+D_\theta)}\Big]    
+\kappa \Delta \Gamma \frac{3v_0^2D_\theta
  e^{-\kappa\bar{\Gamma}t}\sinh{\kappa\bar{\Gamma}t}}
{2\kappa\bar{\Gamma}(\kappa\bar{\Gamma}+D_\theta)(\kappa\bar{\Gamma}-D_\theta)(\kappa\bar{\Gamma}+2D_\theta)}\Bigg]
\end{split}
\label{eq:28c}
   \end{equation}
\end{widetext}

\section{Persistence probability}
\label{sec:per_prob_harm}

We now want to calculate an expression for the persistence probability
of an active anisotropic particle in the presence of a harmonic trap.
As before, we first calculate the two-time correlation function of the
$x$-coordinate of the position vector. Keeping up to the first order
correction in $\kappa \Delta \Gamma/2$, the expression for the two
time correlation function becomes
$\langle x(t_1) x(t_2) \rangle_{\xi, \theta}=\langle x_0(t_1)x_0(t_2)\rangle_{\xi, \theta} -\kappa
\Delta \Gamma \langle x_0(t_1)x_1(t_2)\rangle_{\xi, \theta} $, with $t_1>t_2$. Note
that the quantities $\langle x_0(t_1) x_1(t_2) \rangle$ and
$\langle x_0(t_2) x_1(t_1) \rangle_{\xi, \theta}$ are only equal in the asymptotic
limit, which is also the limit under consideration when evaluating the
persistence probability. In \cref{R_0R_0} and \cref{subsec:R_0R_1} we
have explicitly shown the calculations of the two terms that appear in
the expression for the two-time correlation function
$\langle x(t_1) x(t_2)\rangle_{\xi, \theta}$. We merely quote the final expression
here:
\begin{equation}
    \begin{split}
        &\langle
        x_0(t_1)x_0(t_2)\rangle_{\xi, \theta}=\frac{k_BT}{\kappa}e^{-\kappa\Bar{\Gamma}t_1}\Big[e^{\kappa\Bar{\Gamma}t_2}-e^{-\kappa\Bar{\Gamma}t_2}\Big]\\
        &+\frac{k_BT}{\kappa}(\kappa\Delta\Gamma)\cos{2\theta_0}e^{-\kappa\Bar{\Gamma}t_1}\Big[\frac{e^{(\kappa\Bar{\Gamma}-4D_\theta)t_2}
        -e^{-\kappa\Bar{\Gamma}t_2}}{2\kappa\Bar{\Gamma}-4D_\theta}\Big]\\
      &+\frac{v_0^2\cos{2\theta_0}}{2}\Bigg[\frac{2D_\theta(e^{-\kappa\Bar{\Gamma}t_1}e^{(\kappa\Bar{\Gamma}-4D_\theta)t_2}-
        e^{-\kappa\Bar{\Gamma}(t_1+t_2)})}{(2\kappa\Bar{\Gamma}-4D_\theta)(\kappa\Bar{\Gamma}-D_\theta)(\kappa\Bar{\Gamma}-3D_\theta)}\\
      &+\frac{e^{-D_\theta t_1}e^{-3D_\theta
          t_2}-e^{-\kappa\Bar{\Gamma}t_2}e^{-D_\theta
          t_1}}{(\kappa\Bar{\Gamma}-3D_\theta)(\kappa\Bar{\Gamma}-D_\theta)}\\
           &-\frac{e^{-\kappa\Bar{\Gamma}t_1}e^{-D_\theta t_2}
        -e^{-\kappa\Bar{\Gamma}(t_1+t_2)}}{(\kappa\Bar{\Gamma}-3D_\theta)(\kappa\Bar{\Gamma}-D_\theta)}\Bigg]\\
      &+\frac{v_0^2}{2}\Bigg[\frac{2D_\theta(e^{-\kappa\Bar{\Gamma}(t_1+t_2)}-e^{-\kappa\Bar{\Gamma}(t_1-t_2)})}{2\kappa\Bar{\Gamma}
        (\kappa\Bar{\Gamma}-D_\theta)(\kappa\Bar{\Gamma}+D_\theta)}\\
      &+\frac{e^{-D_\theta
          (t_1-t_2)}-e^{-D_\theta t_1}
        e^{-\kappa\Bar{\Gamma}t_2}}{(\kappa\Bar{\Gamma}-D_\theta)(\kappa\Bar{\Gamma}+D_\theta)}\\
      &-\frac{e^{-\kappa\Bar{\Gamma}t_1}e^{-D_\theta
          t_2}
        -e^{-\kappa\Bar{\Gamma}(t_1+t_2)}}{(\kappa\Bar{\Gamma}-D_\theta)(\kappa\Bar{\Gamma}+D_\theta)}\Bigg]
    \end{split}
    \label{28d}
\end{equation}
and
\begin{equation}
\begin{split}
\langle
x_0(t_1)x_1(t_2)\rangle_{\xi, \theta}&=\Big(\frac{k_BT}{\kappa}\Big)\cos{2\theta_0}e^{-\kappa\overline{\Gamma}t_1}
\bigg(\frac{e^{(\kappa\overline{\Gamma}-4D_\theta)t_2}-e^{-\kappa\overline{\Gamma}t_2}}{2\kappa\overline{\Gamma}-4D_\theta}\\
&-\frac{e^{-\kappa\overline{\Gamma}t_2}-
  e^{-(4D_\theta+\kappa\overline{\Gamma})t_2}}{4D_\theta}\bigg)\\
&+\Big(\frac{k_BT}{\kappa}\Big)\Big(\frac{\Delta\Gamma}{2\overline{\Gamma}}\Big)e^{-\kappa\overline{\Gamma}t_1}
\bigg[\frac{e^{\kappa\overline{\Gamma}t_2}-e^{-\kappa\overline{\Gamma}t_2}}{2\kappa\overline{\Gamma}+4D_\theta}\\
&-\frac{2\kappa\overline{\Gamma}}{4D_\theta}\frac{e^{-\kappa\overline{\Gamma}t_2}-e^{-(\kappa\overline{\Gamma}+4D_\theta)t_2}}
{\kappa\overline{\Gamma}+4D_\theta}\bigg]\\
&-\frac{3v_0^2D_\theta
  e^{-\kappa\bar{\Gamma}t_1}\sinh{\kappa\bar{\Gamma}t_2}}
{2\kappa\bar{\Gamma}(\kappa\bar{\Gamma}+D_\theta)(\kappa\bar{\Gamma}-D_\theta)(\kappa\bar{\Gamma}+2D_\theta)}
\end{split}
\label{28e}
\end{equation}

Choosing initial angle $\theta_0=0$, all the terms in both the
expressions for $\langle x_0(t_1)x_0(t_2)\rangle_{\xi, \theta}$ and
$\langle x_0(t_1)x_1(t_2)\rangle_{\xi, \theta}$ survive. However, since we are
interested in the asymptotic limit with $t_1>>t_2$,  we drop the 
terms which are higher order exponentials in time and therefore decay
faster. Further, since we are interested in the first order
correction, we drop the second bracketed term in \cref{28e}.

After a little algebra, the two-time correlation function for the
$x$-coordinate of the position vector, with initial angle $\theta_0=0$
becomes
\begin{equation}
    \begin{split}
      &\langle x(t_1)x(t_2)\rangle_{\xi,\theta,\theta_0=0}\\
      &=e^{-\kappa\Bar{\Gamma}t_1}\Bigg[\Big(\frac{2k_BT}{\kappa'}\Big)
      \sinh{\kappa\Bar{\Gamma}t_2}
      +\frac{v_0^2(e^{-\kappa\Bar{\Gamma}t_2}-e^{-D_\theta
          t_2})}{(\kappa\Bar{\Gamma}-3D_\theta)(\kappa\Bar{\Gamma}+D_\theta)}\\
      &+\left(\frac{2k_B
      T}{\kappa}\right)\left(\frac{\kappa
      \Delta \Gamma}{2}\right)e^{-\kappa \bar{\Gamma} t_2}\left(\frac{1-e^{-4D_\theta t_2}}{4D_\theta}\right)\Bigg]
    \end{split}
    \label{28f}
  \end{equation}
  where the effective trap constant $\kappa^\prime$ is defined as
  $\kappa^{\prime-1}=\kappa^{-1}[1-\frac{
    v_0^2D_\theta}{\Bar{D}(\kappa\Bar{\Gamma}-D_\theta)(\kappa\Bar{\Gamma}+D_\theta)}+\frac{\kappa
    \Delta
    \Gamma}{2}\frac{3v_0^2D_\theta}{2\bar{D}(\kappa\bar{\Gamma}+D_\theta)(\kappa\bar{\Gamma}-D_\theta)(\kappa\bar{\Gamma}+2D_\theta)}\Big]$
  As before, we define the variable
  $X(t)=x(t)/\sqrt{\langle x^2\rangle_{\xi,\theta}}$ and the
  correlation function of $\langle X(t_1)X(t_2)\rangle_{\xi,\theta}$
  is given by

\onecolumngrid
\vspace*{2.5cm}
\begin{equation}
\begin{split}
  \langle X(t_1)X(t_2)\rangle_{\xi,\theta}=
  \frac{e^{-\kappa\bar{\Gamma}t_1/2}}{e^{-\kappa\bar{\Gamma}t_2/2}}
  \left[\frac{\Big(\frac{2k_BT}{\kappa^\prime}\Big)
      \sinh{\kappa\Bar{\Gamma}t_1}+\frac{v_0^2(e^{-\kappa\Bar{\Gamma}t_1}-e^{-D_\theta
          t_1})}{(\kappa\Bar{\Gamma}-3D_\theta)(\kappa\Bar{\Gamma}+D_\theta)}+\left(\frac{2k_B
      T}{\kappa}\right)\left(\frac{\kappa
      \Delta \Gamma}{2}\right)e^{-\kappa \bar{\Gamma}
    t_1}\left(\frac{1-e^{-4D_\theta
        t_1}}{4D_\theta}\right)}{\Big(\frac{2k_BT}{\kappa^\prime}\Big)\sinh{\kappa\Bar{\Gamma}t_2}
    +\frac{v_0^2(e^{-\kappa\Bar{\Gamma}t_2}-e^{-D_\theta
          t_2})}{(\kappa\Bar{\Gamma}-3D_\theta)(\kappa\Bar{\Gamma}+D_\theta)}
   +\left(\frac{2k_B
      T}{\kappa}\right)\left(\frac{\kappa
      \Delta \Gamma}{2}\right)e^{-\kappa \bar{\Gamma} t_2}\left(\frac{1-e^{-4D_\theta t_2}}{4D_\theta}\right)}\right]^{1/2}
\end{split}
\label{eq:28g}
\end{equation}
\twocolumngrid
Using the time transformation for an imaginary time variable $T$, such that
\begin{equation}
\begin{split}
    e^T&=e^{-\kappa\Bar{\Gamma}t}\Bigg[\Big(\frac{2k_BT}{\kappa^\prime}\Big)\sinh{\kappa\Bar{\Gamma}t}\\
    &   +\frac{v_0^2(e^{-\kappa\Bar{\Gamma}t}-e^{-D_\theta
          t})}{(\kappa\Bar{\Gamma}-3D_\theta)(\kappa\Bar{\Gamma}+D_\theta)}+\left(\frac{2k_B
      T}{\kappa}\right)\left(\frac{\kappa
      \Delta \Gamma}{2}\right)\\
  &e^{-\kappa \bar{\Gamma} t}\frac{1-e^{-4D_\theta t}}{4D_\theta}\Bigg]
\end{split}
    \label{28h}
\end{equation}
the two time correlation function in \cref{eq:28g} transforms into a
stationary correlator of the form $C(T_1-T_2)=e^{-(T_1-T_2)/2}$ and
the persistence probability in the asymptotic limit in the imaginary
variable $T$ is given by $p(T) \sim e^{-T/2}$. Transforming back into
real-time, the persistence probability becomes

\begin{figure}
    \centering
    \includegraphics[width=0.5\textwidth]{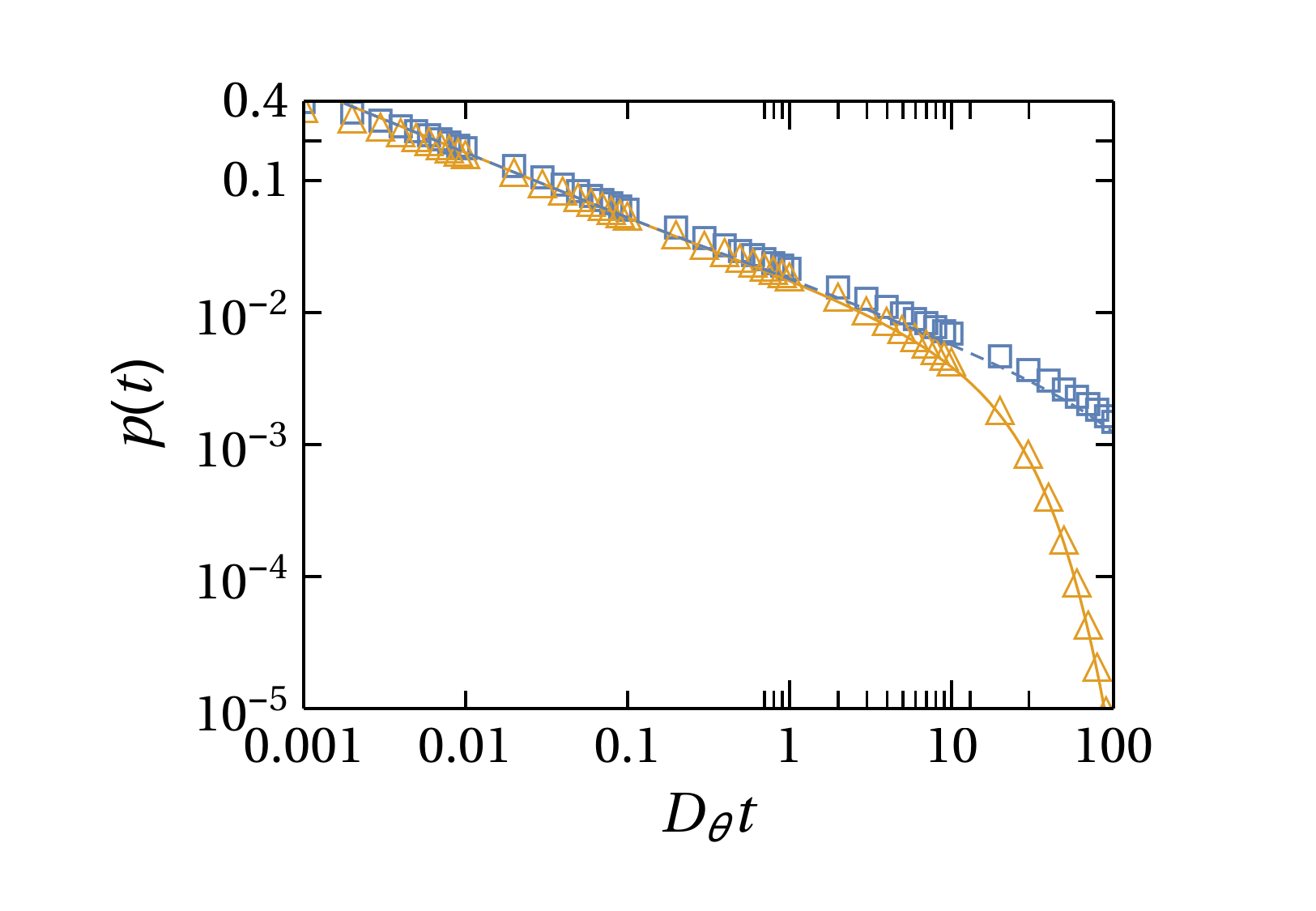}
    \caption{Plot of $p(t)$ for different choices of stiffness of the
      potential $\kappa$ of the harmonically trapped anisotropic
      particle(blue square for $\kappa=0.01$ and orange triangle for
      $\kappa=0.1$) for self-propelled velocity $v_0=0.05$: the
      colours representing different stiffness of the potential are
      written above the plot. The rotational diffusion constant and
      initial angle $\theta_0$ were fixed at $D_\theta=1$ and
      $\theta_0=0$, translational diffusivities are fixed as
      $D_\parallel=1$, $D_\perp=0.5$. The solid lines are plots of
      Eq.(\ref{har}) with the appropriate values of $\kappa$,
      $D_{\parallel}$, $D_{\perp}$, and $D_\theta$.}
    \label{fig100}
\end{figure}

\begin{equation}
    \begin{split}
      p(t,\theta_0=0)&=e^{-\kappa\Bar{\Gamma}t/2}\Bigg[\Big(\frac{2k_BT}{\kappa^\prime}\Big)\sinh{\kappa\Bar{\Gamma}t}\\
      &+\frac{v_0^2(e^{-\kappa\Bar{\Gamma}t}-e^{-D_\theta
          t})}{(\kappa\Bar{\Gamma}-3D_\theta)(\kappa\Bar{\Gamma}+D_\theta)}\\
      &+\left(\frac{2k_B T}{\kappa}\right)\left(\frac{\kappa
      \Delta \Gamma}{2}\right)
  e^{-\kappa \bar{\Gamma} t}\frac{1-e^{-4D_\theta t}}{4D_\theta}\Bigg]^{-1/2}
    \end{split}
    \label{har}
  \end{equation}
  In the limit of $v_0 \to 0$, the equation correctly reproduces the
  result for a passive anisotropic particle. \cite{ghosh2020}. In
  order to validate the equation, we performed numerical simulations
  of \cref{20} with the initial condition chosen from a Gaussian
  distribution with a very small width, so that the sign of
  $\vec{r}(0)$ is well defined. The trajectories were evolved in time
  with an integration time step of $\delta t=0.001$. The persistence
  probability was determined from the fraction of trajectories for
  which $x(t)$ did not change its sign. A comparison of the measured
  persistence probability is shown in \cref{fig100} for two values
  $\kappa$. There is an excellent agreement of the measured survival
  probability with the analytical expression given in \cref{har}.
\section{Conclusion}
In brief, we have calculated the persistence probability along the
$x$-axis of an active anisotropic particle in two dimensions in the
absence of any potential and in the presence of a harmonic
potential. The two-time correlation function has been calculated in
the both cases. In the case of the harmonic trapped particle, we have
used a perturbative solution for calculating the correlation
functions. The persistence probability has been calculated with
suitable space and time transformations. The anaytic expressions has
been validated against numerically measured persistence
probability. 

\bibliographystyle{unsrt}




\onecolumngrid
\appendixpage

\begin{appendices}
\numberwithin{equation}{section}




\section{Mean-square displacement of a active anisotropic  free particle}
\label{sec:free_particle}

\begin{equation}
    \begin{split}
      I_1&=\int_{0}^{t}\langle\cos{\theta}(t^\prime)\cos{\theta}(t^{\prime\prime})\rangle dt^\prime dt^{\prime\prime}\\
      &=\frac{1}{2}\int_{0}^{t}dt^\prime\int_{0}^{t}dt^{\prime\prime}\langle\Big[\cos{(\theta(t^\prime)
        +\theta(t^{\prime\prime}))}+\cos{(\theta(t^\prime)-\theta(t^{\prime\prime}))}\Big]\\
      &=\frac{1}{2}\cos{2\theta_0}\int_{0}^{t}e^{-D_\theta[t^\prime+t^{\prime\prime}+2
        min(t^\prime,t^{\prime\prime})]}dt^\prime
      dt^{\prime\prime}+\frac{1}{2}\int_{0}^{t}e^{-D_\theta[t^\prime+t^{\prime\prime}-2
        min(t^\prime,t^{\prime\prime})]}dt^\prime dt^{\prime\prime}
    \end{split}
    \label{A1}
\end{equation}

Integral $I_1$ is having two separate integrals and solving these two separately 

\begin{equation}
    \begin{split}
    I^\prime &=\int_{0}^{t}e^{-D_\theta[t^\prime+t^{\prime\prime}+2 min(t^\prime,t^{\prime\prime})]}dt^\prime dt^{\prime\prime}\\
    &=\int_{0}^{t}dt^\prime \int_{0}^{t^\prime}dt^{\prime\prime}e^{-D_\theta(t^\prime+3t^{\prime\prime})}+\int_{0}^{t}dt^{\prime\prime}\int_{0}^{t^{\prime\prime}}dt^\prime e^{-D_\theta(3t^\prime+t^{\prime\prime})}\\
    &=\int_{0}^{t}e^{-D_\theta t^\prime}dt^\prime\int_{0}^{t^\prime}e^{-3D_\theta t^{\prime\prime}}dt^{\prime\prime}+\int_{0}^{t}e^{-D_\theta t^{\prime\prime}}dt^{\prime\prime}\int_{0}^{t^{\prime\prime}}e^{-3D_\theta t^\prime}dt^\prime\\
    &=\frac{1}{6D_\theta^2}(3-4e^{-D_\theta t}+e^{-4D_\theta t})
    \end{split}
    \label{A2}
\end{equation}

and 

\begin{equation}
    \begin{split}
        I^{\prime\prime}&=\int_{0}^{t}dt^\prime\int_{0}^{t}dt^{\prime\prime}e^{-D_\theta[t^\prime+t^{\prime\prime}-2 min(t^\prime, t^{\prime\prime})]}\\
        &=\int_{0}^{t}dt^\prime\int_{0}^{t^\prime}dt^{\prime\prime}e^{-D_\theta(t^\prime-t^{\prime\prime})}+\int_{0}^{t}dt^{\prime\prime}\int_{0}^{t^{\prime\prime}}dt^\prime e^{-D_\theta(t^{\prime\prime}-t^\prime)}\\
        &=\int_{0}^{t}e^{-D_\theta t^\prime}dt^\prime\int_{0}^{t^\prime}e^{D_\theta t^{\prime\prime}}dt^{\prime\prime}+\int_{0}^{t}e^{-D_\theta t^{\prime\prime}}dt^{\prime\prime}\int_{0}^{t^{\prime\prime}}e^{D_\theta t^\prime}dt^{\prime}\\
        &=\frac{2}{D_\theta^2}(D_\theta t+e^{-D_\theta t}-1)
    \end{split}
    \label{A3}
\end{equation}

So values of Eq. (\ref{11}) and Eq. (\ref{12}) is added to get $I_1$ as

\begin{equation}
    \begin{split}
    I_1=\frac{\cos{2\theta_0}}{12D_\theta^2}(3-4e^{-D_\theta t}+e^{-4D_\theta t})+\frac{1}{D_\theta^2}(D_\theta t+e^{-D_\theta t}-1)
    \end{split}
    \label{A4}
\end{equation}

Now the second integral of Eq. (\ref{9}) is calculated as

\begin{equation}
    \begin{split}
        I_2&=\int_{0}^{t}dt^\prime\int_{0}^{t}dt^{\prime\prime}\langle\xi_i(t^\prime)\xi_i(t^{\prime\prime})\rangle\\
        &=2k_BT\int_{0}^{t}dt^\prime\int_{0}^{t}dt^{\prime\prime}\langle\Gamma_{ii}[\theta(t^\prime)]\rangle_{\xi_\theta}\delta(t-t^\prime)\\
        &=2k_BT\int_{0}^{t}dt^\prime\langle\Gamma_{ii}[\theta(t^\prime)]\rangle_{\xi_\theta}
    \end{split}
    \label{A5}
\end{equation}

Using the explicit form of $\Gamma_{xx}$ from Eq. (\ref{7}) the
mean-square displacement along the $x$- direction becomes

\begin{equation}
    \begin{split}
      I_2=2k_BT\Big[\Bar{\Gamma}t+\frac{\Delta\Gamma}{2}\cos{2\theta_0}\Big(\frac{1-e^{-4D_\theta
          t}}{4D_\theta}\Big)\Big]
    \end{split}
    \label{A6}
\end{equation}

\section{Calculation of two-time Correlation for a free active
  anisotropic particle}
\label{ssec:two_time_corr_free}

\begin{equation}
    \begin{split}
      I_3&=v_0^2\int
      _{0}^{t_1}dt_1^\prime\int_{0}^{t_2}dt_2^\prime\langle\cos{\theta(t_1^\prime)}
      \cos{\theta(t_2^\prime)}\rangle\\
      &=\frac{v_0^2}{2}\int_{0}^{t_1}dt_1^\prime\int_{0}^{t_2}dt_2^\prime\cos{2\theta_0}e^{-D_\theta[t_1^\prime+t_2^\prime+2
        min(t_1^\prime , t_2^\prime
        )]}+\frac{v_0^2}{2}\int_{0}^{t_1}dt_1^\prime\int_{0}^{t_2}dt_2^\prime
      e^{-D_\theta[t_1^\prime+t_2^\prime-2 min(t_1^\prime , t_2^\prime
        )]}
    \end{split}
    \label{A7}
\end{equation}
for $t_1^\prime<t_2^\prime$
\begin{equation*}
    \begin{split}
        t_1^\prime+t_2^\prime\pm 2t_1^\prime&=3t_1^\prime+t_2^\prime\\
        &=-t_1^\prime+t_2^\prime
    \end{split}
\end{equation*}
for $t_1^\prime>t_2^\prime$
\begin{equation*}
    \begin{split}
        t_1^\prime+t_2^\prime\pm 2t_2^\prime&=t_1^\prime+3t_2^\prime\\
        &=t_1^\prime-t_2^\prime
    \end{split}
\end{equation*}
Lets take the two integrals of Eq.(\ref{A7}) as $I_3^\prime$ and $I_3^{\prime\prime}$. The integral Eq.(\ref{A7}) can be calculated in general terms like,
\begin{equation*}
    \begin{split}
        I_3&=\int_{0}^{t_2}dt_2^\prime\int_{0}^{t_2^\prime} e^{-D_\theta(\alpha t_1^\prime+\beta t_2^\prime)}+\int_{0}^{t_2^\prime}dt^\prime\int_{t_2^\prime}^{t_1}dt_1^\prime e^{-D_\theta(\beta t_1^\prime+\alpha t_2^\prime)}\\
        &=\frac{1-e^{-\beta D_\theta t_2}}{\alpha\beta D_\theta^2}-\frac{1-e^{-(\alpha+\beta)D_\theta t_2}}{\alpha(\alpha+\beta)D_\theta^2}+\frac{1-e^{-(\alpha+\beta)D_\theta t_2}}{\beta(\alpha+\beta)D_\theta^2}-e^{-\beta D_\theta t_1}\frac{1-e^{-\alpha D_\theta t_2}}{\alpha\beta D_\theta^2}
    \end{split}
\end{equation*}
for integral $I_3^\prime$, $(\alpha,\beta)\equiv(3,1)$
\begin{equation*}
    \begin{split}
    I_3^\prime&=\frac{1-e^{-D_\theta t_2}}{6D_\theta^2}+\frac{1-e^{-4D_\theta t_2}}{12D_\theta^2}-e^{-D_\theta t_1}\frac{1-e^{-3D_\theta t_2}}{6D_\theta^2}
    \end{split}
\end{equation*}
and for integral $I_3^{\prime\prime}$, $(\alpha,\beta)\equiv(-1,1)$

\begin{equation*}
    \begin{split}
        I_3^{\prime\prime}=-\frac{1-e^{-D_\theta t_2}}{2D_\theta^2}+\frac{t_2}{D_\theta}-e^{-D_\theta t_1}\frac{1-e^{D_\theta t_2}}{D_\theta^2}
    \end{split}
\end{equation*}

So final form of $I_3$ becomes

\begin{equation}
    \begin{split}
    I_3&=v_0^2\Bigg[\cos{2\theta_0}\Big(\frac{1-e^{-D_\theta t_2}}{6D_\theta^2}+\frac{1-e^{-4D_\theta t_2}}{12D_\theta^2}-e^{-D_\theta t_1}\frac{1-e^{-3D_\theta t_2}}{6D_\theta^2}\Big)-\frac{1-e^{-D_\theta t_2}}{2D_\theta^2}+\frac{t_2}{D_\theta}-e^{-D_\theta t_1}\frac{1-e^{D_\theta t_2}}{D_\theta^2}\Bigg]
    \end{split}
    \label{A10}
\end{equation}

The second integral $I_4$ of \cref{9} is solved as

\begin{equation}
    \begin{split}
    I_4&=\int_{0}^{t_1}dt_1^\prime\int_{0}^{t_2}dt_2^\prime\langle \xi_x(t_1^\prime)\xi_x(t_2^\prime)\rangle\\
    &=2k_BT\Bar{\Gamma}t_2\Big[1+\frac{\Delta\Gamma}{2\Bar{\Gamma}}\cos{2\theta_0}\Big(\frac{1-e^{-4D_\theta t_2}}{4D_\theta t_2}\Big)\Big]
    \end{split}
    \label{A11}
\end{equation}

\section{Calculation of $\langle
       R_{0,i}(t)R_{0,j}(t)\rangle$ for a harmonically
  trapped particle}
\label{R0R0}

\begin{equation}
    \begin{split}
        \langle R_{0,i}(t)R_{0,j}(t)\rangle&=\int_{0}^{t}dt^\prime\int_{0}^{t}dt^{\prime\prime}e^{-\kappa\Bar{\Gamma}(t-t^\prime)}e^{-\kappa\Bar{\Gamma}(t-t^{\prime\prime})}\langle\xi(t^\prime)\xi(t^{\prime\prime})\rangle+\int_{0}^{t}dt^\prime\int_{0}^{t}dt^{\prime\prime}e^{-\kappa\Bar{\Gamma}(t-t^\prime)}e^{-\kappa\Bar{\Gamma}(t-t^{\prime\prime})}
        v_0^2\langle\bold{\hat{n}}(t^\prime)\bold{\hat{n}}(t^{\prime\prime})\rangle
    \end{split}
    \label{A12}
\end{equation}

There are two integrals, lets say $I_5$ and $I_6$ respectively. Lets calculate these two separately 

\begin{equation}
    \begin{split}
      I_5&=\int_{0}^{t}dt^\prime\int_{0}^{t}dt^{\prime\prime}e^{-\kappa\Bar{\Gamma}(t-t^\prime)}
      e^{-\kappa\Bar{\Gamma}(t-t^{\prime\prime})}\langle\xi(t^\prime)\xi(t^{\prime\prime})\rangle\\
      &=\int_{0}^{t}dt^\prime\int_{0}^{t}dt^{\prime\prime}e^{-\kappa\Bar{\Gamma}(t-t^\prime)}
      e^{-\kappa\Bar{\Gamma}(t-t^{\prime\prime})}\Big[\Bar{\Gamma}\mathbf{1}+\frac{\Delta\Gamma}{2}
      \Bar{\Bar{\mathcal{R}}}(t^\prime)\Big]\delta(t^\prime-t^{\prime\prime})\\
      &=2k_BTe^{-2\kappa\Bar{\Gamma}t}\int_{0}^{t}dt^\prime
      e^{2\kappa\Bar{\Gamma}t^\prime}\Big[
      \Bar{\Gamma}\mathbf{1}+\frac{\Delta\Gamma}{2}\langle\Bar{\Bar{\mathcal{R}}}(t^\prime)\rangle\Big]\\
      &=\frac{k_BT}{\kappa}\mathbf{1}(1-e^{-2\kappa\Bar{\Gamma}t})+2k_BTe^{-2\kappa\Bar{\Gamma}t}
      \int_{0}^{t}dt^\prime
      e^{2\kappa\Bar{\Gamma}t^\prime}\frac{\Delta\Gamma}{2}\Bar{\Bar{\mathcal{R}}}(\theta_0)
      e^{-4D_\theta t^\prime}\\
      &=\frac{k_BT}{\kappa}\mathbf{1}(1-e^{-2\kappa\Bar{\Gamma}t})+\Delta
      D\Bar{\Bar{\mathcal{R}}}(\theta_0)e^{-2\kappa\Bar{\Gamma}t}\Big(
      \frac{e^{(2\kappa\Bar{\Gamma}-4D_\theta)}-1}{2\kappa\Bar{\Gamma}-4D_\theta}\Big)
    \end{split}
    \label{A13}
\end{equation}
For $x$-direction
\begin{equation}
    I_5=\frac{k_BT}{\kappa}(1-e^{-2\kappa\Bar{\Gamma}t})+\Delta
    D\cos{2\theta_0}\Big(\frac{e^{-4D_\theta t}
      -e^{-2\kappa\Bar{\Gamma}t}}{2\kappa\Bar{\Gamma}-4D_\theta}\Big)
    \label{A14}
\end{equation}

\begin{equation}
    \begin{split}
      I_6&=\int_{0}^{t}dt^\prime\int_{0}^{t}dt^{\prime\prime}e^{-\kappa\Bar{\Gamma}(t-t^\prime)}
      e^{-\kappa\Bar{\Gamma}(t-t^{\prime\prime})}v_0^2\langle\bold{\hat{n}}(t^\prime)\bold{\hat{n}}(t^{\prime\prime})\rangle\\
      &=v_0^2e^{-2\kappa\Bar{\Gamma}t}\int_{0}^{t}dt^\prime\int_{0}^{t}dt^{\prime\prime}
      e^{\kappa\Bar{\Gamma}(t^\prime+t^{\prime\prime})}\langle\cos{\theta(t^\prime)}\cos{\theta(t^{\prime\prime})}\rangle\\
      &=\frac{v_0^2}{2}e^{-2\kappa\Bar{\Gamma}t}\int_{0}^{t}dt^\prime\int_{0}^{t}dt^{\prime\prime}
      e^{\kappa\Bar{\Gamma}(t^\prime+t^{\prime\prime})}\Big[\cos{2\theta_0}e^{-D_\theta[t^\prime+t^{\prime\prime}+2
        min(t^\prime,
        t^{\prime\prime})]}+e^{-D_\theta[t^\prime+t^{\prime\prime}-2
        min(t^\prime, t^{\prime\prime})]}\Big]
    \end{split}
    \label{A15}
\end{equation}
Lets solve the integrals separately 

\begin{equation}
    \begin{split}
      I_6^\prime&=\frac{v_0^2}{2}e^{-2\kappa\Bar{\Gamma}t}\int_{0}^{t}dt^\prime\int_{0}^{t}dt^{\prime\prime}e^{\kappa\Bar{\Gamma}(t^\prime+t^{\prime\prime})}\cos{2\theta_0}e^{-D_\theta[t^\prime+t^{\prime\prime}+2 min(t^\prime, t^{\prime\prime})]}\\
      &=\frac{v_0^2\cos{2\theta_0}}{2}e^{-2\kappa\Bar{\Gamma}t}\Big[\int_{0}^{t}dt^\prime\int_{0}^{t^\prime}dt^{\prime\prime}e^{\kappa\Bar{\Gamma}(t^\prime+t^{\prime\prime})}e^{-D_\theta(t^\prime+3t^{\prime\prime})}+\int_{0}^{t}dt^\prime\int_{t^\prime}^{t}e^{\kappa\Bar{\Gamma}(t^\prime+t^{\prime\prime})}e^{-D_\theta(3t^\prime+t^{\prime\prime})}\Big]\\
      &=\frac{v_0^2\cos{2\theta_0}}{2}e^{-2\kappa\Bar{\Gamma}t}\Big[\int_{0}^{t}e^{(\kappa\Bar{\Gamma}-D_\theta)t^\prime}dt^\prime\int_{0}^{t^\prime}e^{(\kappa\Bar{\Gamma}-3D_\theta)t^{\prime\prime}}dt^{\prime\prime}+\int_{0}^{t}e^{(\kappa\Bar{\Gamma}-3D_\theta)t^\prime}dt^\prime\int_{t^\prime}^{t}e^{(\kappa\Bar{\Gamma}-D_\theta)t^{\prime\prime}}dt^{\prime\prime}\Big]\\
      &=\frac{v_0^2\cos{2\theta_0}}{2}\Big[\frac{2D_\theta(e^{-4D_\theta
          t}-e^{-2\kappa\Bar{\Gamma}t})}{(2\kappa\Bar{\Gamma}-4D_\theta)(\kappa\Bar{\Gamma}-3D_\theta)(\kappa\Bar{\Gamma}-D_\theta)}+\frac{e^{-4D_\theta
          t}-2e^{-(\kappa\Bar{\Gamma}+D_\theta)t}+e^{-2\kappa\Bar{\Gamma}t}}{(\kappa\Bar{\Gamma}-D_\theta)(\kappa\Bar{\Gamma}-3D_\theta)}\Big]
    \end{split}
    \label{A16}
\end{equation}

\begin{equation}
    \begin{split}
        I_6^{\prime\prime}&=\frac{v_0^2}{2}e^{-2\kappa\Bar{\Gamma}t}\int_{0}^{t}dt^\prime\int_{0}^{t}dt^{\prime\prime}e^{-D_\theta[t^\prime+t^{\prime\prime}-2 min(t^\prime, t^{\prime\prime})]}\\
        &=\frac{v_0^2}{2}e^{-2\kappa\Bar{\Gamma}t}\Big[\int_{0}^{t}dt^\prime\int_{0}^{t^\prime}dt^{\prime\prime}e^{\kappa\Bar{\Gamma}(t^\prime+t^{\prime\prime})}e^{=D_\theta(t^\prime-t^{\prime\prime})}+\int_{0}^{t}dt^\prime\int_{t^\prime}^{t}dt^{\prime\prime}e^{\kappa\Bar{\Gamma}(t^\prime+t^{\prime\prime})}e^{-D_\theta(t^{\prime\prime}-t^\prime)}\Big]\\
        &=\frac{v_0^2}{2}e^{-2\kappa\Bar{\Gamma}t}\Big[\int_{0}^{t}e^{(\kappa\Bar{\Gamma}-D_\theta)t^\prime}dt^\prime\int_{0}^{t^\prime}e^{(\kappa\Bar{\Gamma}+D_\theta)t^{\prime\prime}}dt^{\prime\prime}+\int_{0}^{t}e^{(\kappa\Bar{\Gamma}+D_\theta)t^\prime}dt^\prime\int_{t^\prime}^{t}e^{(\kappa\Bar{\Gamma}-D_\theta)t^{\prime\prime}}dt^{\prime\prime}\Big]\\
        &=\frac{v_0^2}{2}\Big[\frac{1-2e^{-(\kappa\Bar{\Gamma}+D_\theta)t}+e^{-2\kappa\Bar{\Gamma}t}}{(\kappa\Bar{\Gamma}-D_\theta)(\kappa\Bar{\Gamma}+D_\theta)}-\frac{D_\theta(1-e^{-2\kappa\Bar{\Gamma}t})}{\kappa\Bar{\Gamma}(\kappa\Bar{\Gamma}-D_\theta)(\kappa\Bar{\Gamma}+D_\theta)}\Big]
    \end{split}
    \label{A17}
\end{equation}

\section{Calculation of $\langle R_{0,i}(t_1)R_{0,j}(t_2)\rangle$}
\label{R_0R_0}
\begin{equation}
  \langle
  R_{0,i}(t_1)R_{0,j}(t_2)\rangle=\int_{0}^{t_1}dt^\prime\int_{0}^{t_2}dt^{\prime\prime}e^{-\kappa\Bar{\Gamma}(t_1-t_1^\prime)}
  e^{-\kappa\Bar{\Gamma}(t_2-t_2^\prime)}\langle\xi(t_1^\prime)\xi(t_2^\prime)\rangle+v_0^2\int_{0}^{t_1}dt^\prime
  \int_{0}^{t_2}dt^{\prime\prime}e^{-\kappa\Bar{\Gamma}(t_1-t_1^\prime)}e^{-\kappa\Bar{\Gamma}(t_2-t_2^\prime)}
  \langle n_i(t_1^\prime) n_j(t_2^\prime)\rangle
    \label{A24}
\end{equation}

The case of $i=j=1$ corresponds to $\langle x_0(t_1) x_0(t_2) \rangle$
and the explicit form of the correlation becomes

\begin{equation}
  \langle
  R_{0,i}(t_1)R_{0,j}(t_2)\rangle=\int_{0}^{t_1}dt^\prime\int_{0}^{t_2}dt^{\prime\prime}e^{-\kappa\Bar{\Gamma}(t_1-t_1^\prime)}
  e^{-\kappa\Bar{\Gamma}(t_2-t_2^\prime)}\langle\xi(t_1^\prime)\xi(t_2^\prime)\rangle+v_0^2\int_{0}^{t_1}dt^\prime
  \int_{0}^{t_2}dt^{\prime\prime}e^{-\kappa\Bar{\Gamma}(t_1-t_1^\prime)}e^{-\kappa\Bar{\Gamma}(t_2-t_2^\prime)}
  \langle \cos\theta(t_1^\prime) \cos\theta(t_2^\prime)\rangle
    \label{A24a}
\end{equation}

It can be calculated as two separately $I_9$ and $I_{10}$ integrals

\begin{equation}
    \begin{split}
      I_9&=\int_{0}^{t_1}dt^\prime\int_{0}^{t_2}dt^{\prime\prime}e^{-\kappa\Bar{\Gamma}(t_1-t_1^\prime)}
      e^{-\kappa\Bar{\Gamma}(t_2-t_2^\prime)}\langle\xi(t_1^\prime)\xi(t_2^\prime)\rangle\\
        &=2k_BTe^{-\kappa\Bar{\Gamma}(t_1+t_2)}\int_{0}^{t_1}dt_1^\prime\int_{0}^{t_2}dt_2^\prime
        e^{\kappa\Bar{\Gamma}(t_1^\prime+t_2^\prime)}
        \Big[\Bar{\Gamma}\delta_{ij}+\frac{\Delta\Gamma}{2}\langle\mathcal{R}_{ij}(t_1^\prime)\rangle\Big]\delta(t_1^\prime-t_2^\prime)\\
        &=2k_BTe^{-\kappa\Bar{\Gamma}(t_1+t_2)}\int_{0}^{min(t_1,t_2)}e^{2\kappa\Bar{\Gamma}t_1^\prime}dt_1^\prime+k_BT\Delta\Gamma
        e^{-\kappa\Bar{\Gamma}(t_1+t_2)}\int_{0}^{min(t_1,t_2)}e^{2\kappa\Bar{\Gamma}t_1^\prime}\langle\mathcal{R}_{ij}(t_1^\prime)\rangle dt_1^\prime\\
        &=\frac{k_BT}{\kappa}\Big[e^{-\kappa\Bar{\Gamma}(t_1-t_2)}-e^{-\kappa\Bar{\Gamma}(t_1+t_2)}\Big]+\frac{2k_BT}{\kappa}
        (\frac{\kappa\Delta\Gamma}{2})\cos{2\theta_0}e^{-\kappa\Bar{\Gamma}(t_1+t_2)}\int_{0}^{min(t_1,t_2)}
        e^{(2\kappa\Bar{\Gamma}-4D_\theta)t_1^\prime}dt_1^\prime\\
        &=\frac{k_BT}{\kappa}\Big[e^{-\kappa\Bar{\Gamma}(t_1-t_2)}-e^{-\kappa\Bar{\Gamma}(t_1+t_2)}\Big]+\frac{k_BT}{\kappa}(\kappa\Delta\Gamma)
        \cos{2\theta_0}e^{-\kappa\Bar{\Gamma}(t_1+t_2)}\Big[\frac{e^{(2\kappa\Bar{\Gamma}-4D_\theta)t_2}-1}{2\kappa\Bar{\Gamma}-4D_\theta}\Big]\\
        &=\frac{k_BT}{\kappa}e^{-\kappa\Bar{\Gamma}t_1}\Big[e^{\kappa\Bar{\Gamma}t_2}-e^{-\kappa\Bar{\Gamma}t_2}\Big]+\frac{k_BT}{\kappa}(\kappa\Delta\Gamma)
        \cos{2\theta_0}e^{-\kappa\Bar{\Gamma}t_1}\Big[\frac{e^{(\kappa\Bar{\Gamma}-4D_\theta)t_2}-
          e^{-\kappa\Bar{\Gamma}t_2}}{2\kappa\Bar{\Gamma}-4D_\theta}\Big]
    \end{split}
    \label{A25}
\end{equation}

\begin{equation}
    \begin{split}
       I_{10}
       &=v_0^2e^{-\kappa\Bar{\Gamma}(t_1+t_2)}\int_{0}^{t_1}dt_1^\prime\int_{0}^{t_2}dt_2^\prime
       e^{\kappa\Bar{\Gamma}(t_1^\prime+t_2^\prime)}
       \langle\cos{\theta}(t_1^\prime)\cos{\theta}(t_2^\prime)\rangle\\
       &=\frac{v_0^2}{2}e^{-\kappa\Bar{\Gamma}(t_1+t_2)}\int_{0}^{t_1}dt_1^\prime\int_{0}^{t_2}dt_2^\prime
       e^{\kappa\Bar{\Gamma}(t_1^\prime+t_2^\prime)}\Big[\cos{2\theta_0}e^{-D_\theta[t_1^\prime+t_2^\prime+2min(t_1^\prime,t_2^\prime)]}
       +e^{-D_\theta[t_1^\prime+t_2^\prime-2min(t_1^\prime,t_2^\prime)]}\Big]
    \end{split}
    \label{A26}
\end{equation}

Lets calculate integrals separately

\begin{equation}
    \begin{split}
      I_{10}^\prime&=\frac{v_0^2\cos{2\theta_0}}{2}e^{-\kappa\Bar{\Gamma}(t_1+t_2)}\int_{0}^{t_1}dt_1^\prime
      \int_{0}^{t_2}dt_2^\prime e^{\kappa\Bar{\Gamma}(t_1^\prime+t_2^\prime)}e^{-D_\theta[t_1^\prime+t_2^\prime+2min(t_1^\prime,t_2^\prime)]}\\
      &=\frac{v_0^2\cos{2\theta_0}}{2}e^{-\kappa\Bar{\Gamma}(t_1+t_2)}\Bigg[\int_{0}^{t_2}dt_2^\prime
      \int_{0}^{t_2^\prime}dt_1^\prime
      e^{\kappa\Bar{\Gamma}(t_1^\prime+t_2^\prime)}e^{-D_\theta(3t_1^\prime+t_2^\prime)}+\int_{0}^{t_2}dt_2^\prime
      \int_{t_2^\prime}^{t_1}dt_1^\prime e^{\kappa\Bar{\Gamma}(t_1^\prime+t_2^\prime)}e^{-D_\theta(t_1^\prime+3t_2^\prime)}\Bigg]\\
      &=\frac{v_0^2\cos{2\theta_0}}{2}e^{-\kappa\Bar{\Gamma}(t_1+t_2)}\Bigg[\int_{0}^{t_2}e^{(\kappa\Bar{\Gamma}-D_\theta)t_2^\prime}
      dt_2^\prime\int_{0}^{t_2^\prime}e^{(\kappa\Bar{\Gamma}-3D_\theta)t_1^\prime}dt_1^\prime+\int_{0}^{t_2}
      e^{(\kappa\Bar{\Gamma}-3D_\theta)t_2^\prime}dt_2^\prime\int_{t_2^\prime}^{t_1}e^{(\kappa\Bar{\Gamma}-D_\theta)t_1^\prime}dt_1^\prime\Bigg]\\
      &=\frac{v_0^2\cos{2\theta_0}}{2}\Bigg[\frac{2D_\theta(e^{-\kappa\Bar{\Gamma}t_1}e^{(\kappa\Bar{\Gamma}-4D_\theta)t_2}-
        e^{-\kappa\Bar{\Gamma}(t_1+t_2)})}{(2\kappa\Bar{\Gamma}-4D_\theta)(\kappa\Bar{\Gamma}-D_\theta)(\kappa\Bar{\Gamma}-3D_\theta)}
      +\frac{e^{-D_\theta t_1}e^{-3D_\theta t_2}-e^{-\kappa\Bar{\Gamma}t_2}e^{-D_\theta t_1}-e^{-\kappa\Bar{\Gamma}t_1}e^{-D_\theta
        t_2}+e^{-\kappa\Bar{\Gamma}(t_1+t_2)}}{(\kappa\Bar{\Gamma}-3D_\theta)(\kappa\Bar{\Gamma}-D_\theta)}\Bigg]
    \end{split}
    \label{A27}
\end{equation}

\begin{equation}
    \begin{split}
       I_{10}^{\prime\prime}&= \frac{v_0^2}{2}e^{-\kappa\Bar{\Gamma}(t_1+t_2)}\int_{0}^{t_1}dt_1^\prime\int_{0}^{t_2}dt_2^\prime e^{\kappa\Bar{\Gamma}(t_1^\prime+t_2^\prime)}e^{-D_\theta[t_1^\prime+t_2^\prime-2min(t_1^\prime,t_2^\prime)]}\\
       &=\frac{v_0^2}{2}e^{-\kappa\Bar{\Gamma}(t_1+t_2)}\Bigg[\int_{0}^{t_2}dt_1^\prime\int_{0}^{t_2^\prime}dt_2^\prime e^{\kappa\Bar{\Gamma}(t_1^\prime+t_2^\prime)}e^{-D_\theta(t_2^\prime-t_1^\prime)}+\int_{0}^{t_2}dt_2^\prime\int_{t_2^\prime}^{t_1}dt_1^\prime e^{\kappa\Bar{\Gamma}(t_1^\prime+t_2^\prime)}e^{-D_\theta(t_1^\prime-t_2^\prime)}\Bigg]\\
       &=\frac{v_0^2}{2}\Bigg[\frac{2D_\theta(e^{-\kappa\Bar{\Gamma}(t_1+t_2)}-e^{-\kappa\Bar{\Gamma}(t_1-t_2)})}{2\kappa\Bar{\Gamma}(\kappa\Bar{\Gamma}-D_\theta)(\kappa\Bar{\Gamma}+D_\theta)}+\frac{e^{-D_\theta t_1}e^{D_\theta t_2}-e^{-D_\theta t_1}e^{-\kappa\Bar{\Gamma}t_2}-e^{-\kappa\Bar{\Gamma}t_1}e^{-D_\theta t_2}+e^{-\kappa\Bar{\Gamma}(t_1+t_2)}}{(\kappa\Bar{\Gamma}-D_\theta)(\kappa\Bar{\Gamma}+D_\theta)}\Bigg]
     \end{split}
     \label{A28}
\end{equation}

\begin{equation}
    \begin{split}
        &\langle x_0(t_1)x_0(t_2)\rangle=\frac{k_BT}{\kappa}e^{-\kappa\Bar{\Gamma}t_1}\Big[e^{\kappa\Bar{\Gamma}t_2}-e^{-\kappa\Bar{\Gamma}t_2}\Big]+\frac{k_BT}{\kappa}(\kappa\Delta\Gamma)\cos{2\theta_0}e^{-\kappa\Bar{\Gamma}t_1}\Big[\frac{e^{(\kappa\Bar{\Gamma}-4D_\theta)t_2}-e^{-\kappa\Bar{\Gamma}t_2}}{2\kappa\Bar{\Gamma}-4D_\theta}\Big]\\
        &+\frac{v_0^2\cos{2\theta_0}}{2}\Bigg[\frac{2D_\theta(e^{-\kappa\Bar{\Gamma}t_1}e^{(\kappa\Bar{\Gamma}-4D_\theta)t_2}-e^{-\kappa\Bar{\Gamma}(t_1+t_2)})}{(2\kappa\Bar{\Gamma}-4D_\theta)(\kappa\Bar{\Gamma}-D_\theta)(\kappa\Bar{\Gamma}-3D_\theta)}+\frac{e^{-D_\theta t_1}e^{-3D_\theta t_2}-e^{-\kappa\Bar{\Gamma}t_2}e^{-D_\theta t_1}-e^{-\kappa\Bar{\Gamma}t_1}e^{-D_\theta t_2}+e^{-\kappa\Bar{\Gamma}(t_1+t_2)}}{(\kappa\Bar{\Gamma}-3D_\theta)(\kappa\Bar{\Gamma}-D_\theta)}\Bigg]\\
        &+\frac{v_0^2}{2}\Bigg[\frac{2D_\theta(e^{-\kappa\Bar{\Gamma}(t_1+t_2)}-e^{-\kappa\Bar{\Gamma}(t_1-t_2)})}{2\kappa\Bar{\Gamma}(\kappa\Bar{\Gamma}-D_\theta)(\kappa\Bar{\Gamma}+D_\theta)}+\frac{e^{-D_\theta t_1}e^{D_\theta t_2}-e^{-D_\theta t_1}e^{-\kappa\Bar{\Gamma}t_2}-e^{-\kappa\Bar{\Gamma}t_1}e^{-D_\theta t_2}+e^{-\kappa\Bar{\Gamma}(t_1+t_2)}}{(\kappa\Bar{\Gamma}-D_\theta)(\kappa\Bar{\Gamma}+D_\theta)}\Bigg]
    \end{split}
    \label{A29}
\end{equation}

It is quite easy to see that substituting $t_1=t_2=t$ in the above
equation reproduces the result $\langle x_0^2(t) \rangle$ that has
been explicitly calculated in \cref{A14},\cref{A16} and \cref{A17}.

\section{Calculation of $\langle R_{0,i}(t_1)R_{1,j}(t_2)\rangle$}
\label{subsec:R_0R_1}

We now turn our attention to the first order correction that enter the
correlation matrix. Using the formal solution for $R_{1,j}(t)$ we
write down
\begin{equation}
    \begin{split}
        &\langle R_{0,i}(t_1)R_{1,j}(t_2)\rangle=\Bigg\langle
        R_{0,i}(t_1)\int_{0}^{t_2}dt_2^\prime
        e^{-\kappa\Bar{\Gamma}(t_2-t_2^\prime)}
        \sum_k\mathcal{R}_{jk}(t_2^\prime)R_{0,k}(t_2^\prime)\Bigg\rangle\\
        &=\Bigg\langle\int_{0}^{t_2}dt_2^\prime
        e^{-\kappa\Bar{\Gamma}(t_2-t_2^\prime)}\sum_k\mathcal{R}_{jk}(t_2^\prime)R_{0,i}(t_1)
        R_{0,k}(t_2^\prime)\Bigg\rangle\\
        &=\Bigg \langle\int_{0}^{t_2}dt_2^\prime
        e^{-\kappa\Bar{\Gamma}(t_2-t_2^\prime)}\sum_k\mathcal{R}_{jk}(t_2^\prime)\int_{0}^{t_1}dt_1^\prime\int_{0}^{t_2^\prime}dt_2^{\prime\prime}
        e^{-\kappa\Bar{\Gamma}(t_1-t_1^\prime)}e^{-\kappa\Bar{\Gamma}(t_2^\prime-t_2^{\prime\prime})}
        \Big[\xi_i(t_1^\prime)\xi_k(t_2^{\prime\prime})+v_0^2 n_i(t_1^\prime)n_k(t_2^{\prime\prime})\Big]\Bigg\rangle
    \end{split}
    \label{A30}
  \end{equation}

We evaluate the two terms in the integral separately. The first term
involves an average over the thermal fluctuations in the position and
we denote it by $I_{11}$. The explicit calculation of $I_{11}$ becomes  
  
\begin{equation}
    \begin{split}
        I_{11}&=\int_{0}^{t_2}dt_2^\prime
        e^{-\kappa\Bar{\Gamma}(t_2-t_2^\prime)}\Bigg\langle\sum_k\mathcal{R}_{jk}(t_2^\prime)\int_{0}^{t_1}dt_1^\prime
        \int_{0}^{t_2^\prime}dt_2^{\prime\prime}e^{-\kappa\Bar{\Gamma}(t_1-t_1^\prime)}e^{-\kappa\Bar{\Gamma}(t_2^\prime-t_2^{\prime\prime})}
        \xi(t_1^\prime)\xi(t_2^{\prime\prime})\Bigg\rangle\\
        &=\int_{0}^{t_2}dt_2^\prime
        e^{-\kappa\Bar{\Gamma}(t_2-t_2^\prime)}\Bigg\langle\sum_k\mathcal{R}_{jk}(t_2^\prime)
        \Big[\frac{k_BT}{\kappa}\delta_{ki}(e^{-\kappa\Bar{\Gamma}(t_1-t_2^\prime)})
        +k_BT\Delta\Gamma
        e^{-\kappa\Bar{\Gamma}(t_1+t_2^\prime)}\int_{0}^{min(t_1,t_2^\prime)}dt^{\prime\prime}
        e^{2\kappa\Bar{\Gamma}t^{\prime\prime}}\mathcal{R}_{ki}(t^{\prime\prime})\Big]\Bigg\rangle\\
        &=\frac{k_BT}{\kappa}e^{-\kappa\Bar{\Gamma}(t_1+t_2)}\int_{0}^{t_2}dt_2^\prime
        e^{\kappa\Bar{\Gamma}t_2^\prime}
        \langle\mathcal{R}_{ji}(t_2^{\prime\prime})\rangle(e^{\kappa\Bar{\Gamma}t_2^\prime}-e^{-\kappa\Bar{\Gamma}t_2^\prime})\\
        &+k_BT\Delta\Gamma e^{-\kappa\Bar{\Gamma}(t_1+t_2)}\int_{0}^{min(t_1,t_2^\prime)}e^{2\kappa\Bar{\Gamma}t^{\prime\prime}}
        \sum_k\Big\langle\mathcal{R}_{jk}(t_2^\prime)\mathcal{R}_{ki}(t^{\prime\prime})\Big\rangle\\
      \end{split}
         \label{A31}
    \end{equation}

 In order to proceed further with the calculation, We consider the correlation function $\langle x_0(t_1)
  x_1(t_2)\rangle$ so that the above expression becomes 
    \begin{equation}
        \begin{split}
 I_{11} &=\frac{k_BT}{\kappa}e^{-\kappa\Bar{\Gamma}(t_1+t_2)}\mathcal{R}_{ji}(\theta_0)\int_{0}^{t_2}dt_2^\prime
        e^{-4D_\theta t_2^\prime}
        (e^{2\kappa\Bar{\Gamma}t_2^\prime}-1)\\
        &+k_BT\Delta\Gamma
        e^{-\kappa\Bar{\Gamma}(t_1+t_2)}\int_{0}^{t_2}dt_2^\prime\int_{0}^{t_2^\prime}dt^{\prime\prime}
        e^{2\kappa\Bar{\Gamma}t^{\prime\prime}}e^{-4D_\theta(t_2^\prime+t^{\prime\prime}-2\min(t_2^\prime,t^{\prime\prime}))}\\
        &=\Big(\frac{k_BT}{\kappa}\Big)\cos{2\theta_0}e^{-\kappa\overline{\Gamma}t_1}\bigg(\frac{e^{(\kappa\overline{\Gamma}-4D_\theta)t_2}
          -e^{-\kappa\overline{\Gamma}t_2}}{2\kappa\overline{\Gamma}-4D_\theta}-\frac{e^{-\kappa\overline{\Gamma}t_2}-
          e^{-(4D_\theta+\kappa\overline{\Gamma})t_2}}{4D_\theta}\bigg)\\
        &+\Big(\frac{k_BT}{\kappa}\Big)\Big(\frac{\Delta\Gamma}{2\overline{\Gamma}}\Big)e^{-\kappa\overline{\Gamma}t_1}
        \bigg[\frac{e^{\kappa\overline{\Gamma}t_2}-e^{-\kappa\overline{\Gamma}t_2}}{2\kappa\overline{\Gamma}+4D_\theta}-
        \frac{2\kappa\overline{\Gamma}}{4D_\theta}\frac{e^{-\kappa\overline{\Gamma}t_2}-e^{-(\kappa\overline{\Gamma}+4D_\theta)t_2}}
        {\kappa\overline{\Gamma}+4D_\theta}\bigg]
    \end{split}
    \label{A31a}
\end{equation}

The second term of \cref{A30} is denoted by $I_{12}$ and is given as 
\begin{equation}
    \begin{split}
        I_{12}&=v_0^2\int_{0}^{t_2}dt_2^\prime
        e^{-\kappa\Bar{\Gamma}(t_2-t_2^\prime)}\Bigg\langle\sum\mathcal{R}_{jk}(t_2^\prime)\int_{0}^{t_1}dt_1^\prime
        \int_{0}^{t_2^\prime}dt_2^{\prime\prime}e^{-\kappa\Bar{\Gamma}(t_1-t_1^\prime)}e^{-\kappa\Bar{\Gamma}(t_2^\prime-t_2^{\prime\prime})}
        n_i(t_1^\prime) n_j(t_2^{\prime\prime})\Bigg\rangle
    \end{split}
    \label{A32}
\end{equation}
so that for the correlation function $\langle x_0(t_1) x_1(t_2)
\rangle$ \cref{A32} transforms as
\begin{equation}
    \begin{split}
        I_{12}=\int_{0}^{t_2}dt_2^\prime
        e^{-\kappa\Bar{\Gamma}(t_2-t_2^\prime)}\int_{0}^{t_1}dt_1^\prime\int_{0}^{t_2^\prime}dt_2^{\prime\prime}
        e^{-\kappa\Bar{\Gamma}(t_1-t_1^\prime)}e^{-\kappa\Bar{\Gamma}(t_2^\prime-t_2^{\prime\prime})}v_0^2[\langle\cos{2\theta}(t_2^\prime)
        \cos{\theta}(t_1^\prime)\cos{\theta}(t_2^{\prime\prime})\rangle+\langle\sin{2\theta}(t_2^\prime)\cos{\theta}(t_1^\prime)
        \cos{\theta}(t_2^{\prime\prime})\rangle]
    \end{split}
    \label{d500}
\end{equation}
We use the following trignometric identities 
\begin{equation}
    \begin{split}
    &\cos{2\theta_1}\cos{\theta_2}\cos{\theta_3}=\frac{1}{4}[\cos({2\theta_1-\theta_2-\theta_3})+\cos({2\theta_1+\theta_2-\theta_3})+\cos({2\theta_1-\theta_2+\theta_3})+\cos{(2\theta_1+\theta_2+\theta_3)}]\\
    &\sin{2\theta_1}\cos{\theta_2}\cos{\theta_3}=\frac{1}{4}[\sin({2\theta_1-\theta_2-\theta_3})+\sin({2\theta_1+\theta_2+\theta_3})+\sin({2\theta_1-\theta_2+\theta_3})+\sin{(2\theta_1+\theta_2+\theta_3)}]
    \end{split}
  \end{equation}
to rewrite the triple product of the trignometric functions, averaged
over the rotational noise as
\begin{equation}
    \begin{split}
        &(a)\quad \langle\cos({2\theta_1-\theta_2-\theta_3})\rangle=e^{-D_\theta(4t_2^\prime+t_1^\prime+t_2^{\prime\prime}-4\min (t_2^\prime,t_1^\prime)-4\min (t_2^\prime,t_2^{\prime\prime})+2\min (t_1^\prime,t_2^{\prime\prime}))}\\
        &(b)\quad \langle\cos({2\theta_1+\theta_2-\theta_3})\rangle=\cos{2\theta_0}e^{-D_\theta(4t_2^\prime+t_1^\prime+t_2^{\prime\prime}+4\min (t_2^\prime,t_1^\prime)-4\min (t_2^\prime,t_2^{\prime\prime})-2\min (t_1^\prime,t_2^{\prime\prime}))}\\
        &(c)\quad \langle\cos({2\theta_1-\theta_2+\theta_3})\rangle=\cos{2\theta_0}e^{-D_\theta(4t_2^\prime+t_1^\prime+t_2^{\prime\prime}-4\min (t_2^\prime,t_1^\prime)+4\min (t_2^\prime,t_2^{\prime\prime})-2\min (t_1^\prime,t_2^{\prime\prime}))}\\
        &(d)\quad \langle\cos({2\theta_1+\theta_2+\theta_3})\rangle=\cos{4\theta_0}e^{-D_\theta(4t_2^\prime+t_1^\prime+t_2^{\prime\prime}+4\min (t_2^\prime,t_1^\prime)+4\min (t_2^\prime,t_2^{\prime\prime})+2\min (t_1^\prime,t_2^{\prime\prime}))}\\
        &(e)\quad \langle\sin({2\theta_1-\theta_2-\theta_3})\rangle=e^{-D_\theta(4t_2^\prime+t_1^\prime+t_2^{\prime\prime}-4\min (t_2^\prime,t_1^\prime)-4\min (t_2^\prime,t_2^{\prime\prime})+2\min (t_1^\prime,t_2^{\prime\prime}))}\\
        &(f) \quad \langle\sin({2\theta_1+\theta_2-\theta_3})\rangle=\sin{2\theta_0}e^{-D_\theta(4t_2^\prime+t_1^\prime+t_2^{\prime\prime}+4\min (t_2^\prime,t_1^\prime)-4\min(t_2^\prime,t_2^{\prime\prime})-2\min (t_1^\prime,t_2^{\prime\prime}))}\\
        &(g)\quad \langle\sin({2\theta_1-\theta_2+\theta_3})\rangle=\sin{2\theta_0}e^{-D_\theta(4t_2^\prime+t_1^\prime+t_2^{\prime\prime}-4\min (t_2^\prime,t_1^\prime)+4\min (t_2^\prime,t_2^{\prime\prime})-2\min (t_1^\prime,t_2^{\prime\prime}))}\\
        &(h)\quad \langle\sin({2\theta_1+\theta_2+\theta_3})\rangle=\sin{4\theta_0}e^{-D_\theta(4t_2^\prime+t_1^\prime+t_2^{\prime\prime}+4\min (t_2^\prime,t_1^\prime)+4\min (t_2^\prime,t_2^{\prime\prime})+2\min (t_1^\prime,t_2^{\prime\prime}))}
    \end{split}
    \label{c0}
  \end{equation}

  Note that in the above set of equations, the time dependence in the
  right hand side of the set of equations (a)--(d) is identical to the
  set of equations (e)--(h).  Let us calculate Eq.(\ref{A31}) by using
  Eq.(\ref{c0}), at first we take the first term
  (a),$\langle\cos({2\theta_1-\theta_2-\theta_3})\rangle$ and calculate
  separately
\begin{equation}
    \begin{split}
      &\frac{v_0^2e^{-\kappa\bar{\Gamma}(t_1+t_2)}}{4}\int_{0}^{t_2}dt_2^\prime\int_{0}^{t_1}dt_1^\prime\int_{0}^{t_2^{\prime}}
      dt_2^{\prime\prime}e^{\kappa\bar{\Gamma}t_1^\prime}e^{\kappa\bar{\Gamma}t_2^{\prime\prime}}
      e^{-D_\theta(4t_2^\prime+t_1^\prime+t_2^{\prime\prime}-4\min (t_2^\prime,t_1^\prime)-4\min (t_2^\prime,t_2^{\prime\prime})+2\min (t_1^\prime,t_2^{\prime\prime}))}
    \end{split}
    \label{c1}
\end{equation}
Here in the above integral always $t_2^\prime>t_2^{\prime\prime}$ and in the first case let us take $t_1^\prime>t_2^\prime$ we get 

Case(1), $t_1^\prime>t_2^\prime$
\begin{equation}
    \begin{split}
        &\frac{v_0^2e^{-\kappa\bar{\Gamma}(t_1+t_2)}}{4}\int_{0}^{t_2}dt_2^\prime\int_{t_2^\prime}^{t_1}dt_1^\prime\int_{0}^{t_2^\prime}dt_2^{\prime\prime}e^{\kappa\bar{\Gamma}t_1^\prime}e^{\kappa\bar{\Gamma}t_2^{\prime\prime}}e^{-4D_\theta t_2^\prime}e^{-D_\theta t_1^\prime}e^{-D_\theta t_2^{\prime\prime}}e^{4D_\theta t_2^\prime}e^{4D_\theta t_2^{\prime\prime}}e^{-2D_\theta t_2^{\prime\prime}}\\
        &=\frac{v_0^2e^{-D_\theta t_1}(e^{D_\theta t_2}-e^{-\kappa\bar{\Gamma}t_2})}{4(\kappa\bar{\Gamma}+D_\theta)^2(\kappa\bar{\Gamma}-D_\theta)}-\frac{v_0^2e^{-\kappa\bar{\Gamma}t_1}(e^{\kappa\bar{\Gamma}t_2}-e^{-\kappa\bar{\Gamma}t_2})}{8\kappa\bar{\Gamma}(\kappa\bar{\Gamma}+D_\theta)(\kappa\bar{\Gamma}-D_\theta)}-\frac{t_2v_0^2e^{-\kappa\bar{\Gamma}t_2}e^{-D_\theta t_1}}{4(\kappa\bar{\Gamma}+D_\theta)(\kappa\bar{\Gamma}-D_\theta)}+\frac{v_0^2e^{-\kappa\bar{\Gamma}t_1}(e^{-D_\theta t_2}-e^{-\kappa\bar{\Gamma}t_2})}{4(\kappa\bar{\Gamma}+D_\theta)(\kappa\bar{\Gamma}-D_\theta)^2}
    \end{split}
    \label{c3}
\end{equation}
Case(2), $t_1^\prime<t_2^\prime$
\begin{equation}
\begin{split}
&\frac{v_0^2e^{-\kappa\bar{\Gamma}(t_1+t_2)}}{4}\Big[\int_{0}^{t_2}dt_2^\prime\int_{0}^{t_2^\prime}dt_1^\prime\int_{0}^{t_1^\prime}dt_2^{\prime\prime}e^{\kappa\bar{\Gamma}t_1^\prime}e^{\kappa\bar{\Gamma}t_2^{\prime\prime}}e^{-4D_\theta t_2^\prime}e^{3D_\theta t_1^\prime}e^{3D_\theta t_2^{\prime\prime}}e^{-2D_\theta t_2^{\prime\prime}}\\
&+\int_{0}^{t_2}dt_2^\prime\int_{0}^{t_2^\prime}dt_1^\prime\int_{t_1^\prime}^{t_2^\prime}dt_2^{\prime\prime}e^{\kappa\bar{\Gamma}t_1^\prime}e^{\kappa\bar{\Gamma}t_2^{\prime\prime}}e^{-4D_\theta t_2^\prime}e^{3D_\theta t_1^\prime}e^{3D_\theta t_2^{\prime\prime}}e^{-2D_\theta t_1^\prime}\Big]\\
&=\frac{v_0^2e^{-\kappa\bar{\Gamma}t_1}}{8(\kappa\bar{\Gamma}+D_\theta)(\kappa\bar{\Gamma}+2D_\theta)}\Big[\frac{\sinh{\kappa\bar{\Gamma}t_2}}{\kappa\bar{\Gamma}}-\frac{e^{-\kappa\bar{\Gamma}t_2}-e^{-(\kappa\bar{\Gamma}+4D_\theta)t_2}}{4D_\theta}\Big]-\frac{v_0^2e^{-\kappa\bar{\Gamma}t_1}}{4(\kappa\bar{\Gamma}+D_\theta)(\kappa\bar{\Gamma}+3D_\theta)}\Big[\frac{e^{-D_\theta t_2}-e^{-\kappa\bar{\Gamma}t_2}}{\kappa\bar{\Gamma}-D_\theta}\\
&-\frac{e^{-\kappa\bar{\Gamma}t_2}-e^{-(\kappa\bar{\Gamma}+4D_\theta)t_2}}{4D_\theta}\Big]+\frac{v_0^2e^{-\kappa\bar{\Gamma}t_1}}{4(\kappa\bar{\Gamma}+D_\theta)(\kappa\bar{\Gamma}+3D_\theta)}\Big[\frac{\sinh{\kappa\bar{\Gamma}t_2}}{\kappa\bar{\Gamma}}-\frac{e^{-D_\theta t_2}-e^{-\kappa\bar{\Gamma}t_2}}{\kappa\bar{\Gamma}-D_\theta}\Big]-\frac{v_0^2e^{-\kappa\bar{\Gamma}t_1}}{8(\kappa\bar{\Gamma}+2D_\theta)(\kappa\bar{\Gamma}+3D_\theta)}\Big[\frac{\sinh{\kappa\bar{\Gamma}t_2}}{\kappa\bar{\Gamma}}\\
&-\frac{e^{-\kappa\bar{\Gamma}t_2}-e^{-(\kappa\bar{\Gamma}+4D_\theta)t_2}}{4D_\theta}\Big]
\end{split}
\label{c4}
\end{equation}

Adding Eq.(\ref{c3}) and Eq.(\ref{c4}) terms we get the term Eq.(\ref{c1}) as,
\begin{equation}
\begin{split}
I_a&=\frac{v_0^2e^{-D_\theta t_1}(e^{D_\theta t_2}-e^{-\kappa\bar{\Gamma}t_2})}{4(\kappa\bar{\Gamma}+D_\theta)^2(\kappa\bar{\Gamma}-D_\theta)}+\frac{t_2v_0^2e^{-\kappa\bar{\Gamma}t_2}e^{-D_\theta t_1}}{4(\kappa\bar{\Gamma}+D_\theta)(\kappa\bar{\Gamma}-D_\theta)}-\frac{3v_0^2D_\theta e^{-\kappa\bar{\Gamma}t_1}\sinh{\kappa\bar{\Gamma}t_2}}{4\kappa\bar{\Gamma}(\kappa\bar{\Gamma}+D_\theta)(\kappa\bar{\Gamma}-D_\theta)(\kappa\bar{\Gamma}+2D_\theta)}\\
&-\frac{v_0^2e^{-\kappa\bar{\Gamma}t_1}(e^{-\kappa\bar{\Gamma}t_2}-e^{-(\kappa\bar{\Gamma}+4D_\theta)t_2})}{16(\kappa\bar{\Gamma}+2D_\theta)(\kappa\bar{\Gamma}+D_\theta)(\kappa\bar{\Gamma}+3D_\theta)}-\frac{v_0^2e^{-\kappa\bar{\Gamma}t_1}(\kappa\bar{\Gamma}-5D_\theta)(e^{-D_\theta t_2}-e^{-\kappa\bar{\Gamma}t_2})}{4(\kappa\bar{\Gamma}+D_\theta)(\kappa\bar{\Gamma}+3D_\theta)(\kappa\bar{\Gamma}-D_\theta)^2}
\end{split}
\label{Ia}
\end{equation}
The way first term has been calculated similiarly other terms are calculated to find the exact expression of Eq.(\ref{d500}). The results of Integrals due to terms (b), (c), and (d) are as follows
\begin{equation}
    \begin{split}
        I_b&=\frac{v_0^2e^{-\kappa\bar{\Gamma}t_1}}{8\kappa\bar{\Gamma}(\kappa\bar{\Gamma}+5D_\theta)}\Big[\frac{e^{(\kappa\bar{\Gamma}-4D_\theta)t_2}-e^{-\kappa\bar{\Gamma}t_2}}{2\kappa\bar{\Gamma}-4D_\theta}-\frac{e^{-\kappa\bar{\Gamma}t_2}-e^{-(\kappa\bar{\Gamma}+4D_\theta)t_2}}{4D_\theta}\Big]-\frac{v_0^2e^{-\kappa\bar{\Gamma}t_1}}{4(\kappa\bar{\Gamma}+5D_\theta)(\kappa\bar{\Gamma}-5D_\theta)}\Big[\frac{e^{-9D_\theta t_2}-e^{-\kappa\bar{\Gamma}t_2}}{\kappa\bar{\Gamma}-9D_\theta}\\
        &-\frac{e^{-\kappa\bar{\Gamma}t_2}-e^{-(\kappa\bar{\Gamma}+4D_\theta)t_2}}{4D_\theta}\Big]+\frac{v_0^2e^{-\kappa\bar{\Gamma}t_1}}{4(\kappa\bar{\Gamma}+3D_\theta)(\kappa\bar{\Gamma}-3D_\theta)}\Big[\frac{e^{(\kappa\bar{\Gamma}-4D_\theta)t_2}-e^{-\kappa\bar{\Gamma}t_2}}{2\kappa\bar{\Gamma}-4D_\theta}+\frac{e^{-\kappa\bar{\Gamma}t_2}-e^{-D_\theta t_2}}{\kappa\bar{\Gamma}-D_\theta}\Big]-\frac{v_0^2e^{-\kappa\bar{\Gamma}t_2}}{8\kappa\bar{\Gamma}(\kappa\bar{\Gamma}+3D_\theta)}\\
        &\Big[\frac{e^{(\kappa\bar{\Gamma}-4D_\theta)t_2}-e^{-\kappa\bar{\Gamma}t_2}}{2\kappa\bar{\Gamma}-4D_\theta}-\frac{e^{-\kappa\bar{\Gamma}t_2}-e^{-(\kappa\bar{\Gamma}+4D_\theta)t_2}}{4D_\theta}\Big]+\frac{v_0^2e^{-\kappa\bar{\Gamma}t_1}}{4(\kappa\bar{\Gamma}-D_\theta)(\kappa\bar{\Gamma}+5D_\theta)}\Big[\frac{e^{-9D_\theta t_2}-e^{-\kappa\bar{\Gamma}t_2}}{\kappa\bar{\Gamma}-9D_\theta}-\frac{e^{(\kappa\bar{\Gamma}-4D_\theta)t_2}-e^{-\kappa\bar{\Gamma}t_2}}{2\kappa\bar{\Gamma}-4D_\theta}\Big]\\
        &+\frac{v_0^2e^{-D_\theta t_1}}{4(\kappa\bar{\Gamma}-D_\theta)(\kappa\bar{\Gamma}+5D_\theta)}\Big[\frac{e^{-3D_\theta t_2}-e^{-\kappa\bar{\Gamma}t_2}}{\kappa\bar{\Gamma}-3D_\theta}-\frac{e^{-\kappa\bar{\Gamma}t_2}-e^{-(\kappa\bar{\Gamma}+8D_\theta)t_2}}{8D_\theta}\Big]
    \end{split}
\end{equation}

\begin{equation}
    \begin{split}
        Ic&=\frac{v_0^2e^{-D_\theta t_1}}{4(\kappa\bar{\Gamma}-D_\theta)(\kappa\bar{\Gamma}-3D_\theta)}\Big[\frac{e^{-3D_\theta t_2}-e^{-\kappa\bar{\Gamma}t_2}}{(\kappa\bar{\Gamma}-3D_\theta)}-t_2e^{-\kappa\bar{\Gamma}t_2}\Big]+\frac{v_0^2e^{-\kappa\bar{\Gamma}t_1}}{4(\kappa\bar{\Gamma}-D_\theta)(\kappa\bar{\Gamma}-3D_\theta)}\Big[\frac{e^{-D_\theta t_2}-e^{-\kappa\bar{\Gamma}t_2}}{\kappa\bar{\Gamma}-D_\theta}\\
        &-\frac{e^{(\kappa\bar{\Gamma}-4D_\theta)t_2}-e^{-\kappa\bar{\Gamma}t_2}}{\kappa\bar{\Gamma}-2D_\theta}\Big]+\frac{v_0^2e^{-\kappa\bar{\Gamma}t_1}}{4(\kappa\bar{\Gamma}+5D_\theta)(\kappa\bar{\Gamma}-5D_\theta)}\Big[\frac{e^{-4D_\theta t_2}-e^{-\kappa\bar{\Gamma}t_2}}{\kappa\bar{\Gamma}-4D_\theta}-\frac{e^{-9D_\theta t_2}-e^{-\kappa\bar{\Gamma}t_2}}{\kappa\bar{\Gamma}-9D_\theta}\Big]\\
        &+\frac{v_0^2e^{-\kappa\bar{\Gamma}t_1}}{16\kappa\bar{\Gamma}(\kappa\bar{\Gamma}-3D_\theta)}\Big[\frac{e^{(\kappa\bar{\Gamma}-4D_\theta)t_2}-e^{-\kappa\bar{\Gamma}t_2}}{\kappa\bar{\Gamma}-2D_\theta}-\frac{e^{-\kappa\bar{\Gamma}t_2}-e^{-(\kappa\bar{\Gamma}+4D_\theta)t_2}}{2D_\theta}\Big]+\frac{v_0^2e^{-\kappa\bar{\Gamma}t_1}}{4(\kappa\bar{\Gamma}-3D_\theta)(\kappa\bar{\Gamma}+3D_\theta)}\Big[\frac{e^{-\kappa\bar{\Gamma}t_2}-e^{-(\kappa\bar{\Gamma}+4D_\theta)t_2}}{4D_\theta}\\
        &-\frac{e^{-D_\theta t_2}-e^{-\kappa\bar{\Gamma}t_2}}{\kappa\bar{\Gamma}-D_\theta}\Big]
    \end{split}
\end{equation}

\begin{equation}
\begin{split}
        I_d&=\frac{v_0^2e^{-D_\theta t_1}}{4(\kappa\bar{\Gamma}-D_\theta)(\kappa\bar{\Gamma}-7D_\theta)}\Big[\frac{e^{-15D_\theta t_2}-e^{-\kappa\bar{\Gamma}t_2}}{\kappa\bar{\Gamma}-15D_\theta}-\frac{e^{-\kappa\bar{\Gamma}t_2}-e^{-(\kappa\bar{\Gamma}+8D_\theta)t_2}}{8D_\theta}\Big]+\frac{v_0^2e^{-\kappa\bar{\Gamma}t_1}}{4(\kappa\bar{\Gamma}-D_\theta)(\kappa\bar{\Gamma}-7D_\theta)}\Big[\frac{e^{-9D_\theta t_2}-e^{-\kappa\bar{\Gamma}t_2}}{\kappa\bar{\Gamma}-9D_\theta}\\
        &-\frac{e^{(\kappa\bar{\Gamma}-16D_\theta)t_2}-e^{-\kappa\bar{\Gamma}t_2}}{2(\kappa\bar{\Gamma}-8D_\theta)}\Big]+\frac{v_0^2e^{-\kappa\bar{\Gamma}t_1}}{8(\kappa\bar{\Gamma}-7D_\theta)(\kappa\bar{\Gamma}-6D_\theta)}\Big[\frac{e^{(\kappa\bar{\Gamma}-16D_\theta)t_2}-e^{-\kappa\bar{\Gamma}t_2}}{2\kappa\bar{\Gamma}-16D_\theta}-\frac{e^{-\kappa\bar{\Gamma}t_2}-e^{-(\kappa\bar{\Gamma}+4D_\theta)t_2}}{4D_\theta}\Big]\\
        &+\frac{v_0^2e^{-\kappa\bar{\Gamma}t_1}}{4(\kappa\bar{\Gamma}-5D_\theta)(\kappa\bar{\Gamma}-7D_\theta)}\Big[\frac{e^{-\kappa\bar{\Gamma}t_2}-e^{-(\kappa\bar{\Gamma}+4D_\theta)t_2}}{4D_\theta}-\frac{e^{-9D_\theta t_2}-e^{-\kappa\bar{\Gamma}t_2}}{\kappa\bar{\Gamma}-9D_\theta}\Big]+\frac{v_0^2e^{-\kappa\bar{\Gamma}t_1}}{4(\kappa\bar{\Gamma}-5D_\theta)(\kappa\bar{\Gamma}-7D_\theta)}\Big[\frac{e^{(\kappa\bar{\Gamma}-16D_\theta)t_2}-e^{-\kappa\bar{\Gamma}t_2}}{2\kappa\bar{\Gamma}-16D_\theta}\\
        &-\frac{e^{-9D_\theta t_2}-e^{-\kappa\bar{\Gamma}t_2}}{\kappa\bar{\Gamma}-9D_\theta}\Big]+\frac{v_0^2e^{-\kappa\bar{\Gamma}t_1}}{8(\kappa\bar{\Gamma}-5D_\theta)(\kappa\bar{\Gamma}-6D_\theta)}\Big[\frac{e^{-\kappa\bar{\Gamma}t_2}-e^{-(\kappa\bar{\Gamma}+4D_\theta)t_2}}{4D_\theta}-\frac{e^{(\kappa\bar{\Gamma}-16D_\theta)t_2}-e^{-\kappa\bar{\Gamma}t_2}}{2\kappa\bar{\Gamma}-16D_\theta}\Big]
\end{split}
\end{equation}

Integral values to due term (e) will be same of (a) similarly
others. Terms due to (f), (g), (h) of Eq.(\ref{c0}) will be zero for
the initial orientational angle $\theta_0=0$. Now for the
simplification we are taking only the term associated with
$\sinh{\kappa\bar{\Gamma}t_2}$ of Eq.(\ref{Ia}). Similarly another
same term of $\sinh{\kappa\bar{\Gamma}t_2}$ will arise due to the
contribution of (e) of Eq.(\ref{c0}).

After adding the relevant terms, the final expression for the
two-time correlation term becomes,

\begin{equation}
\begin{split}
\langle
x_0(t_1)x_1(t_2)\rangle&=\Big(\frac{k_BT}{\kappa}\Big)\cos{2\theta_0}e^{-\kappa\overline{\Gamma}t_1}
\bigg(\frac{e^{(\kappa\overline{\Gamma}-4D_\theta)t_2}-e^{-\kappa\overline{\Gamma}t_2}}{2\kappa\overline{\Gamma}-4D_\theta}-\frac{e^{-\kappa\overline{\Gamma}t_2}-
  e^{-(4D_\theta+\kappa\overline{\Gamma})t_2}}{4D_\theta}\bigg)\\
&+\Big(\frac{k_BT}{\kappa}\Big)\Big(\frac{\Delta\Gamma}{2\overline{\Gamma}}\Big)e^{-\kappa\overline{\Gamma}t_1}
\bigg[\frac{e^{\kappa\overline{\Gamma}t_2}-e^{-\kappa\overline{\Gamma}t_2}}{2\kappa\overline{\Gamma}+4D_\theta}-
\frac{2\kappa\overline{\Gamma}}{4D_\theta}\frac{e^{-\kappa\overline{\Gamma}t_2}-e^{-(\kappa\overline{\Gamma}+4D_\theta)t_2}}
{\kappa\overline{\Gamma}+4D_\theta}\bigg]-\frac{3v_0^2D_\theta
  e^{-\kappa\bar{\Gamma}t_1}\sinh{\kappa\bar{\Gamma}t_2}}
{2\kappa\bar{\Gamma}(\kappa\bar{\Gamma}+D_\theta)(\kappa\bar{\Gamma}-D_\theta)(\kappa\bar{\Gamma}+2D_\theta)}
\end{split}
\label{A35}
\end{equation}

From the above equation, the first order correction in $\langle x^2(t) \rangle$ is given by
\begin{equation}
\begin{split}
\langle
x_0(t)x_1(t)\rangle&=\Big(\frac{k_BT}{\kappa}\Big)\cos{2\theta_0}
\bigg(\frac{e^{-4D_\theta t}-e^{-2\kappa\overline{\Gamma}t}}{2\kappa\overline{\Gamma}-4D_\theta}-\frac{e^{-2\kappa\overline{\Gamma}t}-
  e^{-2(2D_\theta+\kappa\overline{\Gamma})t}}{4D_\theta}\bigg)\\
&+\Big(\frac{k_BT}{\kappa}\Big)\Big(\frac{\Delta\Gamma}{2\overline{\Gamma}}\Big)
\bigg[\frac{1-e^{-2\kappa\overline{\Gamma}t}}{2\kappa\overline{\Gamma}+4D_\theta}-
\frac{2\kappa\overline{\Gamma}}{4D_\theta}\frac{e^{-2\kappa\overline{\Gamma}t}-e^{-(2\kappa\overline{\Gamma}+4D_\theta)t}}
{\kappa\overline{\Gamma}+4D_\theta}\bigg]-\frac{3v_0^2D_\theta
  e^{-\kappa\bar{\Gamma}t}\sinh{\kappa\bar{\Gamma}t}}
{2\kappa\bar{\Gamma}(\kappa\bar{\Gamma}+D_\theta)(\kappa\bar{\Gamma}-D_\theta)(\kappa\bar{\Gamma}+2D_\theta)}
\end{split}
\label{A35a}
\end{equation}

\end{appendices}

\end{document}